\documentclass[a4paper,11pt]{article}
\usepackage{jinstpub} % for details on the use of the package, please
\usepackage{gensymb}
\usepackage[modulo]{lineno}

\title{\boldmath The design and construction of the MICE Electron-Muon Ranger}

\author[a]{R. Asfandiyarov,}
\author[a]{P. Bene,}
\author[a]{A. Blondel,}
\author[b,c]{D. Bolognini,}
\author[a]{F. Cadoux,}
\author[a]{S. Debieux,}
\author[a]{F. Drielsma,}
\author[d,e]{G. Giannini,}
\author[a]{J. S. Graulich,}
\author[a]{C. Husi,}
\author[a,1]{Y. Karadzhov,\note{Corresponding author.}}
\author[b,c]{D. Lietti,}
\author[a]{F. Masciocchi,}
\author[a]{L. Nicola,}
\author[a]{E. Noah Messomo,}
\author[b,c]{M. Prest,}
\author[a]{K. Rothenfusser,}
\author[a]{R. Sandstrom,}
\author[e]{E. Vallazza,}
\author[a]{V. Verguilov,}
\author[a]{H. Wisting.}

\affiliation[a]{DPNC, Section de Physique, Universit\'{e} de Gen\`{e}ve, Geneva, Switzerland}
\affiliation[b]{Universit\`{a} degli Studi dell'Insubria, Via Valleggio 11, 22100 Como, Italy}
\affiliation[c]{INFN Milano Bicocca, Piazza della Scienza 3, 20126 Milano, Italy}
\affiliation[d]{Universit\`{a} degli Studi di Trieste, Via A.Valerio, 34127 Trieste, Italy}
\affiliation[e]{INFN Trieste, Padriciano 99, 34012 Trieste, Italy}

\emailAdd{yordan.karadzhov@cern.ch}

\abstract{The Electron-Muon Ranger (EMR) is a fully-active tracking-calorimeter installed in the beam line of the Muon Ionization
Cooling Experiment (MICE). The experiment will demonstrate ionization cooling, an essential technology needed
for the realization of a Neutrino Factory and/or a Muon Collider. The EMR is designed to measure the properties of low
energy beams composed of muons, electrons and pions, and perform the identification particle-by-particle. The detector
consists of 48 orthogonal layers of 59 triangular scintillator bars. The readout is implemented using FPGA custom made
electronics and commercially available modules. This article describes the construction of the detector from its design
up to its commissioning with cosmic data.}

\keywords{Calorimeters, Particle tracking detectors, Muon spectrometers, Front-end electronics for detector readout.}

\begin{document}
\maketitle
\flushbottom

\section{Introduction}
\subsection{Ionization cooling}
The Neutrino Factory~\cite{NFReference, nf} based on a high-energy muon storage-ring is the ultimate tool to study the neutrino mixing
matrix and is established as the best facility to discover, and study the possible leptonic CP violation. It will produce the most intense,
pure and focused neutrino beam ever achieved and is also the first step towards a $\mu^+ \mu^-$ collider~\cite{Neuffer:multitev, Palmer:1996gs}.
The Neutrino Factory accelerator design uses muons as a neutrino source. A proton beam bombards a target to produce pions. These pions are
captured and focused in a high-field solenoid channel and decay to muons, creating a low energy muon beam with very large emittance.  The
emittance (occupied phase-space volume) of the muons needs to be reduced, i.e the muon beam must be \textgravedbl cooled\textacutedbl, so
that the beam can be accelerated efficiently.

\begin{figure}[htb]
 \includegraphics[width=1.\linewidth]{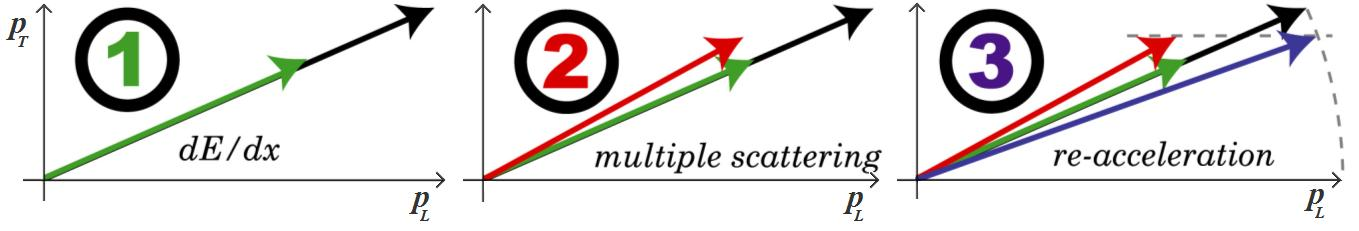}
 \caption{Ionization cooling chronology: (1) energy loss by ionization ($dE/dx$ reduces $p_L$ and $p_T$),
    (2) heating through multiple scattering and
    (3) $p_L$ restored by RF cavities.
   }
 \label{fig:icool}     
\end{figure}

Ionization cooling~\cite{icool1}, summarised in figure~\ref{fig:icool}, provides the only practical solution to muon cooling, because
it is fast enough to cool the beam within the short muon lifetime ($\tau_\mu \sim 2.2 \ \mu s$). The emittance reduction is achieved by
passing muons through a low-Z material (absorber), in which they lose energy via ionization, reducing both their longitudinal and transverse
momentum. The longitudinal component is restored by accelerating cavities, providing a net reduction of the beam emittance. To maximize cooling,
the absorber must be placed at a position where the transverse momentum $p_T$ has a maximum, i.e. at a minimum of the transverse betatron
function, $\beta_{\perp}$.

\subsection{MICE}
The international Muon Ionization Cooling Experiment (MICE) \cite{MICEweb} is under development at the Rutherford Appleton Laboratory (UK).
The goal of the experiment is to build a section of a cooling channel that can demonstrate the principle of ionization cooling.

Since energy loss and multiple Coulomb scattering are momentum dependent, so is the ionization cooling effect. MICE uses a variety of muon
beams of limited intensity and central momenta in the range 140--240\,MeV/$c$ with a spread of $\sim$\,20\,MeV/$c$. These muon beams are
generated using a titanium target \cite{target} dipped into the ISIS proton beam \cite{isis}. Secondary and tertiary particles are captured,
momentum-selected and transported to the cooling section by a system of magnets: a 5\,T superconducting decay solenoid, two dipoles, nine
quadrupoles and a mechanism for inflation of the initial emittance, the diffuser.

\begin{figure}[h]
 \includegraphics[width=.95\textwidth]{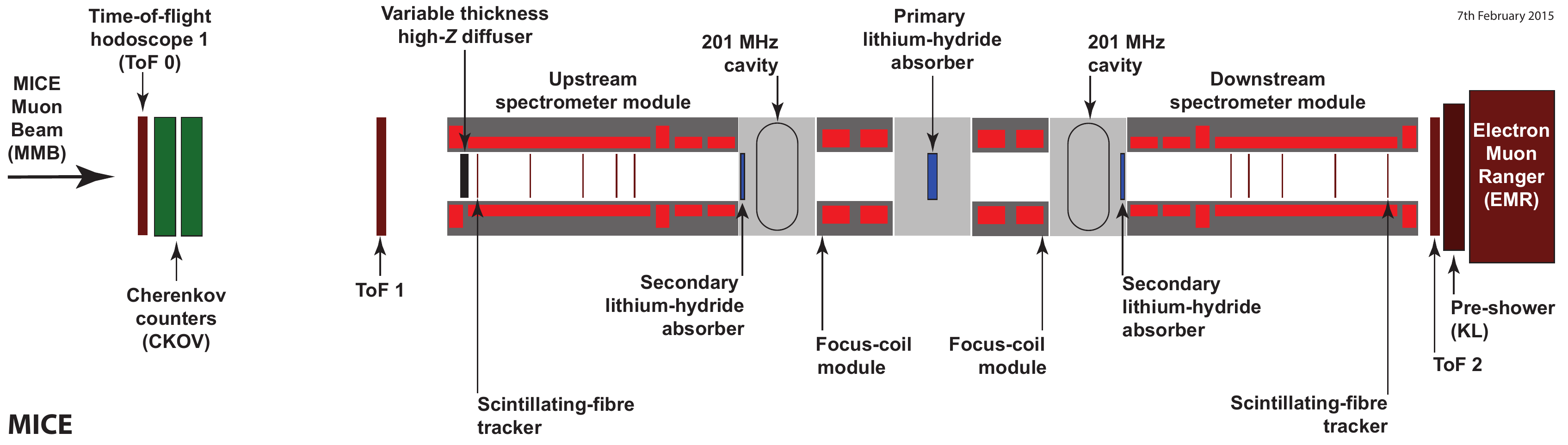}
 % Cooling-demo-labels .pdf: 976x277 pixel, 72dpi, 34.43x9.77 cm, bb=0 0 976 277
 \caption{Schematic view of the MICE experiment.}
 \label{fig:mice}
\end{figure}

The cooling section of MICE is similar to the one of the International Design Study for the Neutrino Factory and is schematically represented
in figure~\ref{fig:mice}. It consists of one primary lithium hydride (LiH) absorber, either side of which are a focus coil, two 201 MHz RF
cavities and two secondary absorbers. The two superconducting focus-coil modules provide strong focusing at the absorber, ensuring that the
transverse betatron function is minimised at this position and enhancing the cooling effect. Each particle is detected individually by two
identical Scintillating fibre trackers in 4\,T solenoids, situated upstream and downstream of the cooling section. The beam emittance is
reconstructed by measuring the position and momentum ($x,y,p_x,p_y,p_z$) of each muon. The two secondary absorbers provide additional cooling
effect, but also serve as shields, protecting the two trackers from the dark-current-induced radiation coming from the RF cavities.

The particle content of the beam is measured by a dedicated system of detectors situated upstream and downstream of the cooling channel and
designed to provide precise muon, pion and electron identification. The upstream part includes two time-of-flight hodoscopes (TOF0 and
TOF1~\cite{TOFref}) and a Cherenkov detector (Ckov~\cite{MICE_PID}). The downstream part combines another time-of-flight hodoscope
(TOF2~\cite{Bertoni:tof2}) with a calorimeter system. The calorimeter system consists of the KLOE-Light (KL~\cite{MICE_PID}) lead-scintillator
sampling calorimeter, operating as a preshower for the totally-active Electron-Muon Ranger (EMR~\cite{ruslan}), placed behind it.

MICE will measure the normalised transverse emittance $\epsilon_N$ with a precision of $\sigma_{\epsilon_N}/\epsilon_N \sim 0.1 \%$ and is
expected to witness a $\sim\,$6\,\% cooling for a muon beam with a nominal momentum of 200\,MeV/$c$ and a 4D input normalised emittance of
$\epsilon_N = 5.8$\,$\pi\cdot$mm$\cdot$rad.

\subsection{EMR}
The EMR is a fully-active scintillator detector. It can be classified as tracking-calorimeter as its granularity allows for track
reconstruction. The primary purpose of the detector is to distinguish muons from their decay products, rejecting events in which
the muon decayed in-flight in the cooling section.  This allows for the selection of a muon beam with a contamination
below 1\,\% \cite{Sandstrom:2007zz}. The range of the muon track can be measured, providing an estimate of the momentum of the muon. 

The construction of the detector started in the early 2011 and in October 2013 the detector was fully commissioned with beam
during one month of dedicated data-taking at RAL. The detector was upgraded in October 2014, including the replacement of the single-anode
photo-multiplier tubes and the installation of a new high-voltage power supply.

\section{EMR Design Concept}
\begin{figure}
 \centering
 \includegraphics[width=0.9\textwidth]{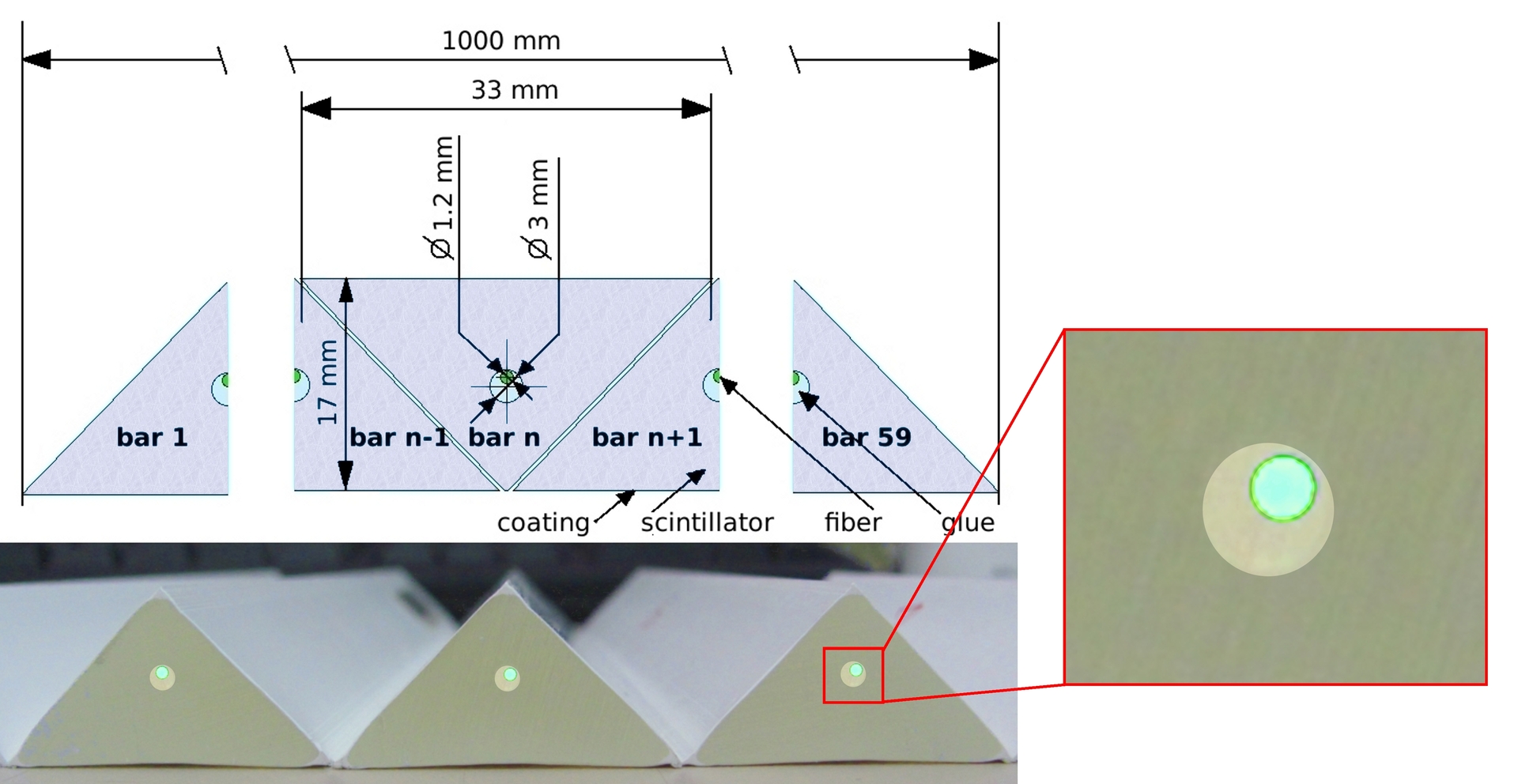}
 \caption[EMR bar cross-section and plane arrangement]{EMR bar cross-section and their arrangement in a plane.}
 \label{fig:bar_arrangement_in_a_plane}
\end{figure}

The EMR consists of triangular scintillator bars arranged in planes. One plane contains 59 bars and covers an area of 1.27~m$^2$. Each even
bar is rotated by 180 degrees with respect to the odd one. A cross-section of bars and their arrangement in a plane is shown in
figure~\ref{fig:bar_arrangement_in_a_plane}. This configuration does not leave dead area for particles crossing a plane with angles
less than 45 degrees with respect to the beam axis.

\begin{figure}
 \centering
 \includegraphics[width=0.9\textwidth]{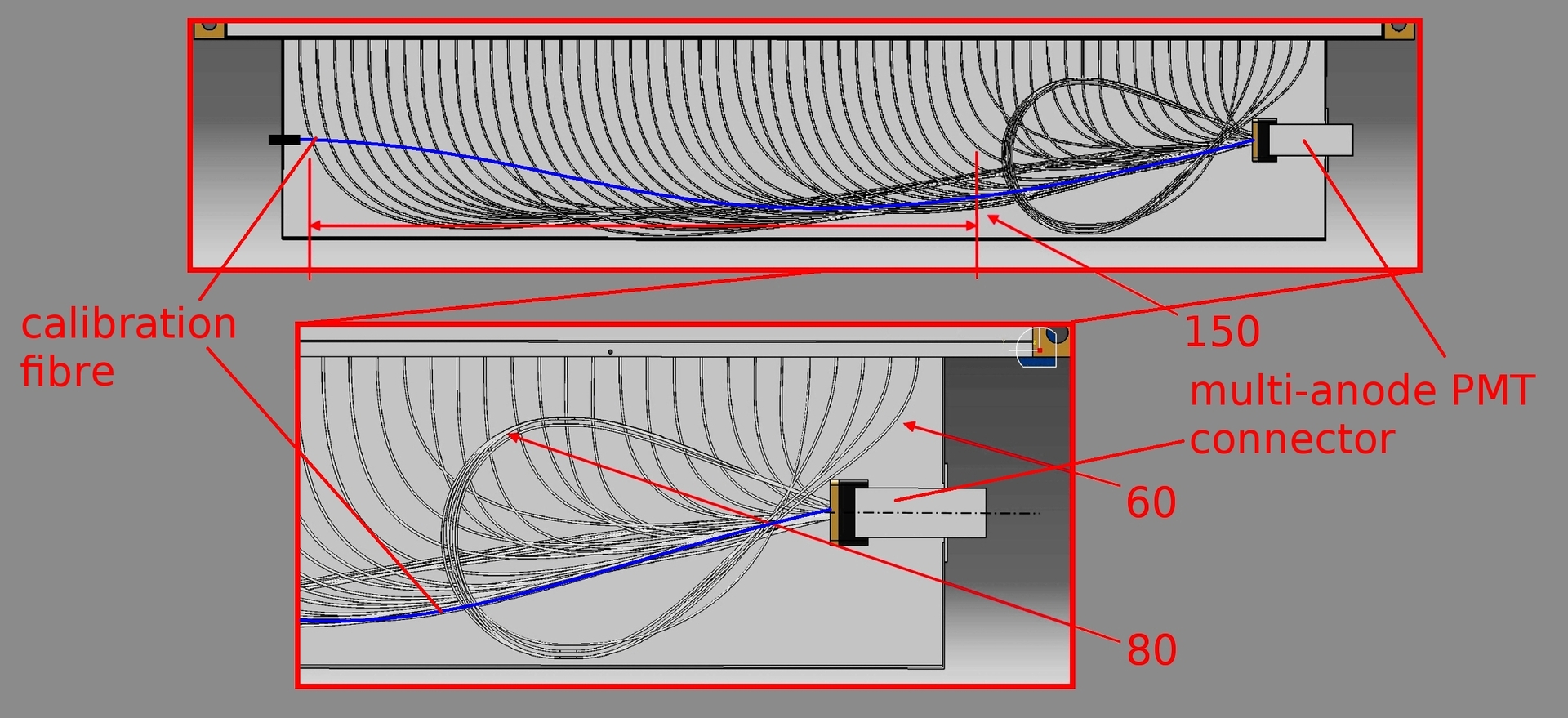}
 \caption[A package of clear fibres in a fibre box]{Clear fibre arrangement in a fibre box on the  multi-anode PMT side. The bending radii of
 a selected few fibres are indicate in red.}
 \label{fig:clear_fibre_package}
\end{figure}

\begin{figure}[htp!]
 \centering
 \includegraphics[width=0.85\textwidth]{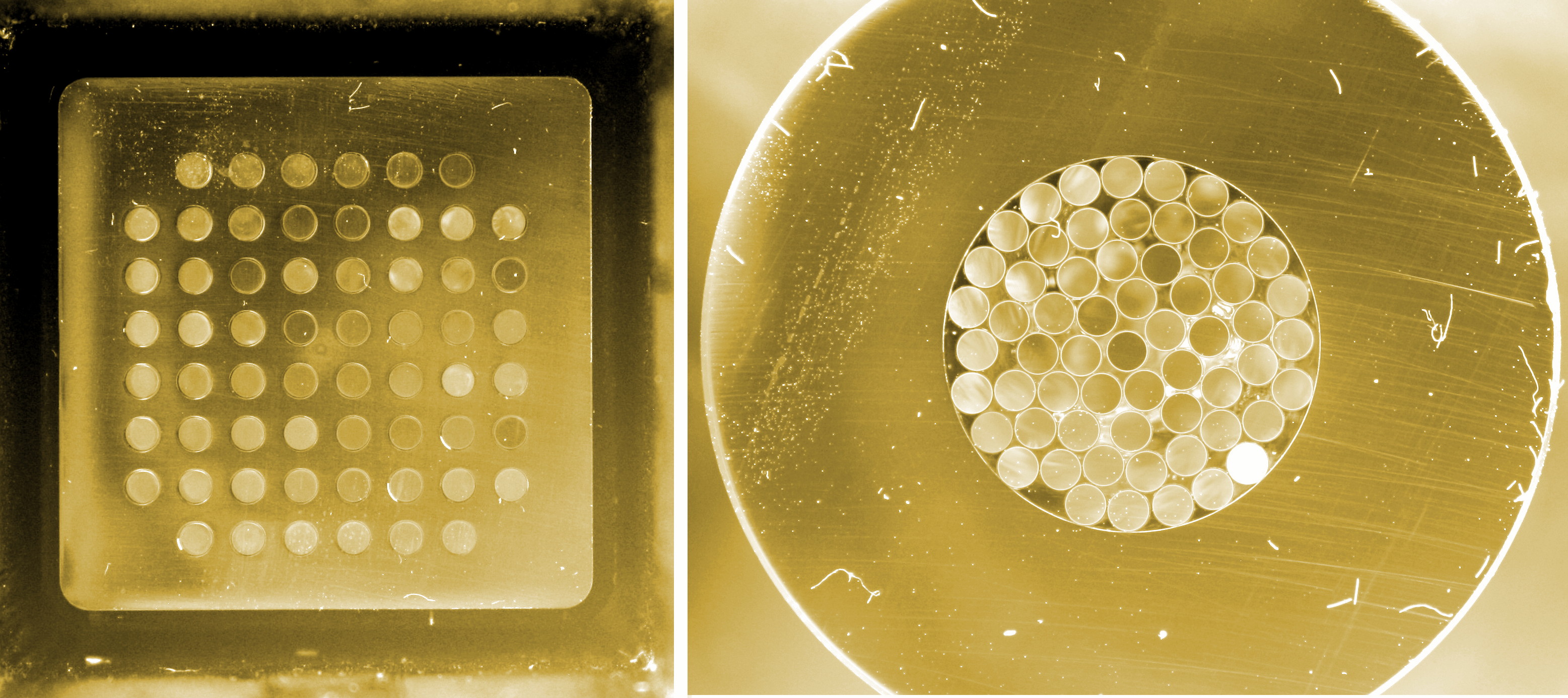}
 \caption[PMT connectors]{Multi-anode PMT connector (left) and single-anode PMT connector (right).}
 \label{fig:pmt_connectors}
\end{figure}

The light, produced when a particle crosses a bar, is collected by a wave-length shifting (WLS) fibre glued inside the bar. At both ends, the
WLS fibre is coupled to clear fibres that transport the light to a photomultiplier tube (PMT). The clear fibres are protected with rubber sleeves
and packed in aluminium fibre boxes as drawn in figure~\ref{fig:clear_fibre_package}. In order to increase the bending radius, which affects light
attenuation, each fibre has its own length. The two bunches of clear fibres coming from the two sides of a plane are glued into different
types of connectors (figure~\ref{fig:pmt_connectors}): one is designed to interface with a multi-anode photomultiplier tube (MAPMT) and the other
with a single-anode photomultiplier tube (SAPMT).

\begin{figure}[htb]
 \centering
 \includegraphics[width=0.85\textwidth]{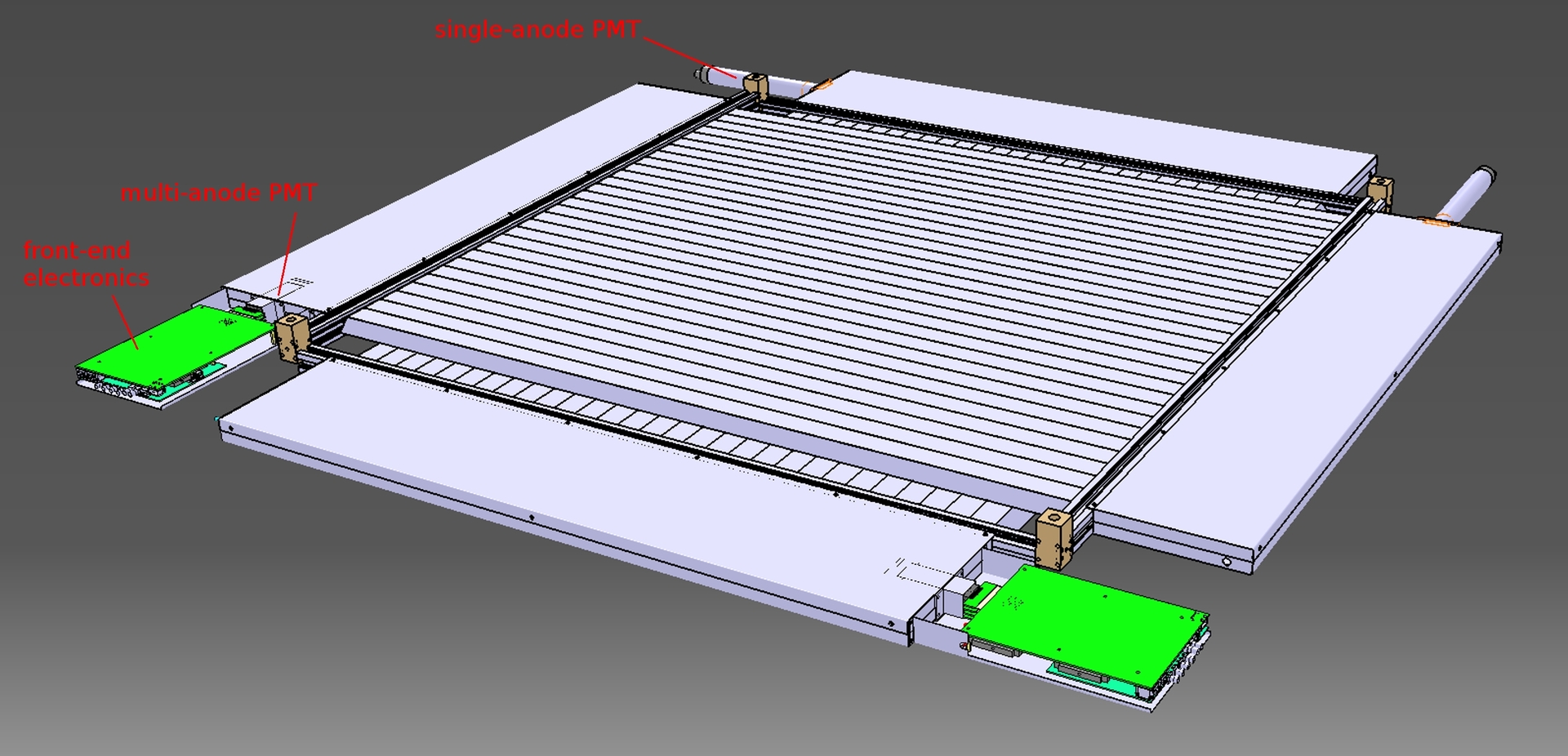}
 \caption[CAD drawing of one EMR module]{CAD drawing of one EMR module made of an X plane and a Y plane. There are two front-end boards
 and two single-anode PMTs per module.}
 \label{fig:emr_module}
\end{figure}

\begin{figure}[htp!]
 \centering
 \includegraphics[width=0.75\textwidth]{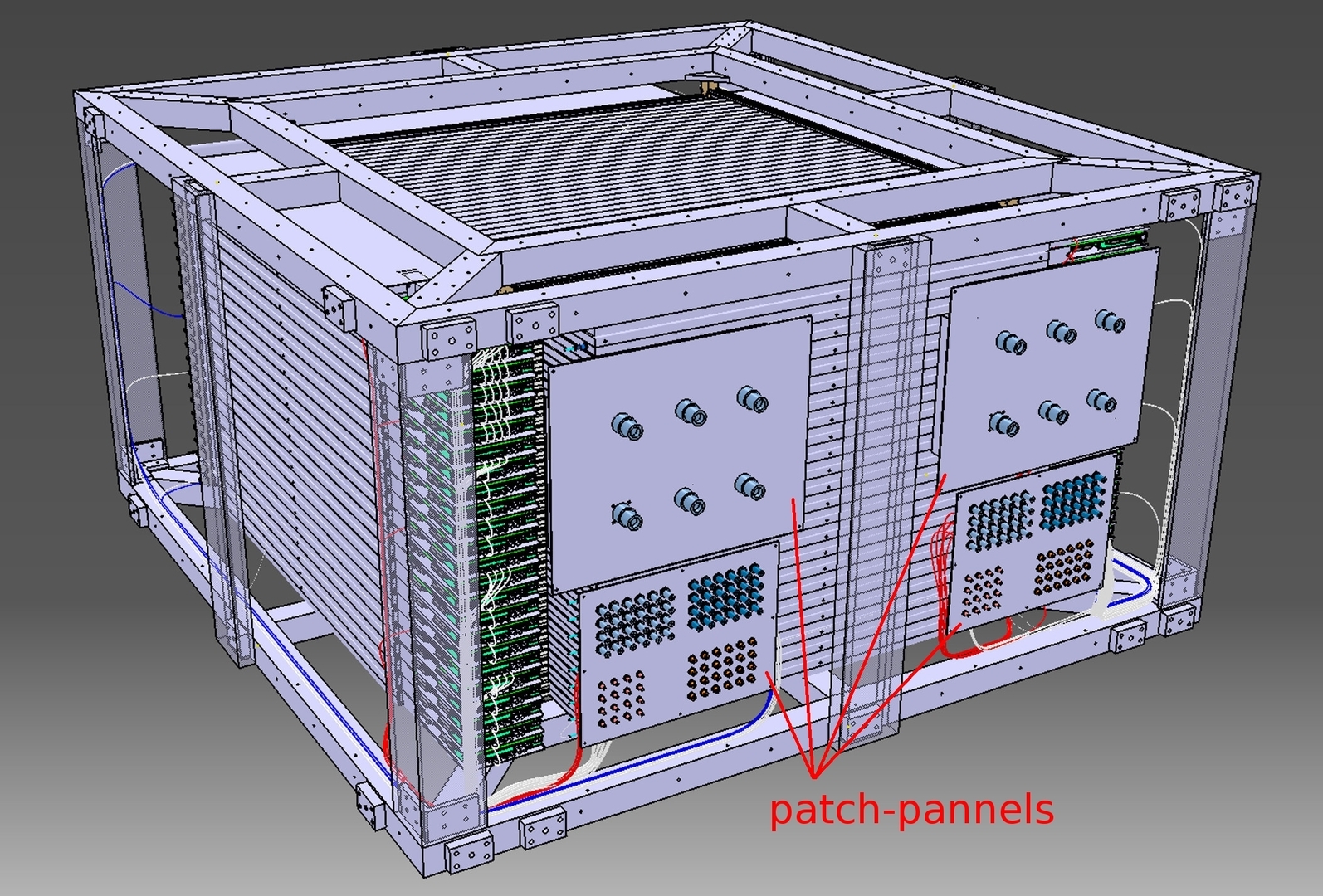}
 \caption[CAD drawing of the EMR detector]{CAD drawing of the EMR detector. The external protective panels are not shown.}
 \label{fig:emr_full_cad_model}
\end{figure}

Two planes attached to each other via aluminum profiles form a rigid structure called a module (figure~\ref{fig:emr_module}). The full detector
contains 24 modules as shown in figure~\ref{fig:emr_full_cad_model}. Panels cover all sides of the detector in order to ensure light-tightness.
The signals coming from a MAPMT are read out and processed by a dedicated front-end board attached directly to the fibre box as shown in
figure~\ref{fig:emr_module}. The SAPMTs are equipped with a voltage divider and the analog signal is sent outside the detector for digitization.

A calibration system was installed inside the enclosure of the detector in order to monitor the drift of the gain and the quantum
efficiency of the PMTs. This system is made of a LED driver distributing light homogeneously to 100 fibres. Each fibre box is
connected to one of the calibration fibres through a connector. Inside a fibre box a clear fibre connects the calibration
fibre to the PMT (see figure~\ref{fig:clear_fibre_package}).

All the cables inside the detector are fed through four patch-panels. There are 96 high-voltage, 6 low-voltage, 48 analog, 48 digital,
and three configuration cables in total. A support frame is designed to withhold the full weight of the sensitive volume ($\sim$\,1\,tonne) with
its readout system. In order to protect the front-end electronics from the magnetic field of the spectrometer solenoid,  a shielding plate is mounted
on the side of the detector that faces the beam. The total weight of the detector is almost 2.5\,tonnes.

\subsection{Optical Elements}
The scintillator bars were manufactured at an extrusion facility at Fermilab~\cite{PlaDalmau:2001fr}. Each bar is 110\,cm long, has a triangular,
1.7\,cm high, 3.3\,cm wide section and is pierced with a 3\,mm hole to host a wavelength shifting fibre. The scintillator is made of polystyrene
pellets\footnote{Dow Styron 663 W} as base, 1\% PPO\footnote{Scintillator, 2,5-diphenyloxazole, C$_{15}$H$_{11}$NO} as primary and
0.03\% POPOP\footnote{Wavelength shifter, 1,4-di-(5-phenyl-2-oxazolyl)-benzene, C$_{24}$H$_{16}$N$_{2}$O, spectrum peaks at 410 nm (violet)} as
secondary fluor. Each bar is coated with TiO$_2$ reflector in order to increase the trapping efficiency. The light output of the scintillator was
measured~\cite{PlaDalmau:2005df} with a PMT (25\% quantum efficiency) and it is around 18 photo-electrons. 

\begin{figure}
 \centering
 \includegraphics[width=.62\textwidth]{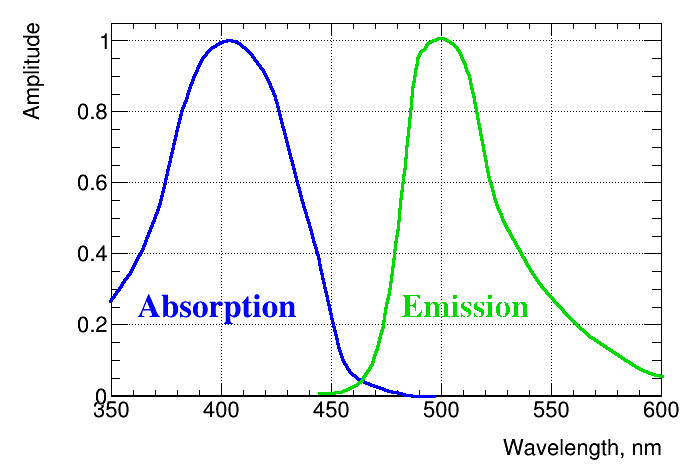}
 \includegraphics[width=.8\textwidth]{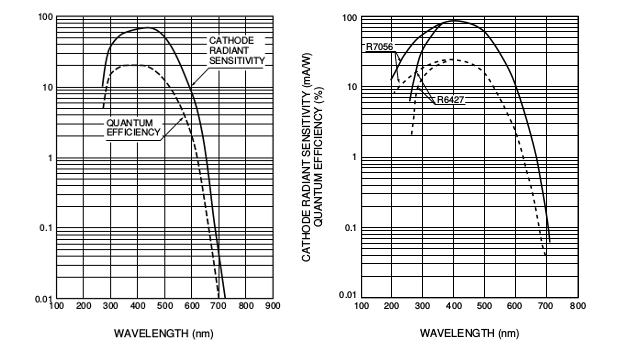} 
 % wls_fiber_emission_spectrum.png: 696x472 pixel, 72dpi, 24.55x16.65 cm, bb=0 0 696 472
 \caption{Top: emission and absorption spectra of the wavelength shifting fiber. Bottom-left: Hamamatsu R5900-00-M64 MAPMT spectral characteristics.
 Bottom-right: Hamamatsu R6427 SAPMT spectral characteristics.}
 \label{qe}
\end{figure}

The WLS fibre glued inside each bar is a double-cladding 1.2\,mm fibre produced by Saint-Gobain Crystals~\cite{saintgobain}.
The core material is polystyrene with acrylic cladding. It has a larger numerical aperture of 0.58 compared to the 0.2--0.3
of graded-index multi-mode fibres used in data communication. Their trapping efficiency is 3.5\,\%. The light is absorbed in the blue part of
the visible spectrum and re-emitted in green (figure~\ref{qe}, top). 

The clear fibres, used to transfer light from the ends of scintillator bar to the PMTs, are 1.5\,mm multi-cladding fibres produced by 
Kuraray~\cite{kuraray} with a special structure (S-type) that allows for better flexibility. The aperture of this fibre matches
the one of WLS fibre to ensure a minimal transmission loss.

\begin{figure}[htp!]
 \centering
 \includegraphics[width=0.8\textwidth]{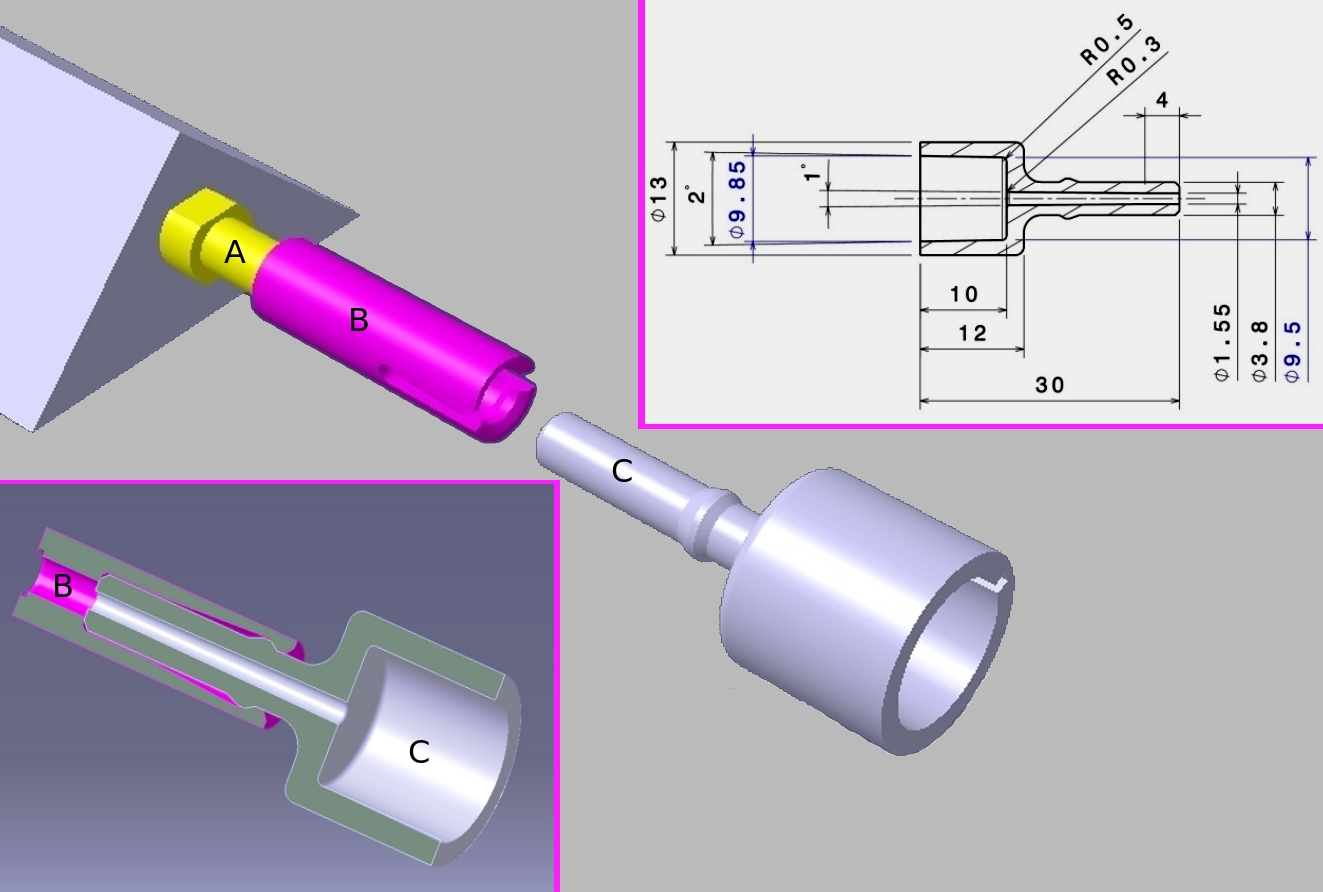}
 \caption[Clear fibre connector]{WLS to clear fibre coupling: a clear fibre connector (C) is attached to the wavelength shifting fibre connector
 (A) using a retaining clip (B).}
 \label{fig:fibre_connectors_cad}
\end{figure}

A special connector was designed to couple the clear fibre to the WLS fibre~\cite{emr_design_change} and is represented in 
figure~\ref{fig:fibre_connectors_cad}. It has a small cylindrical enlargement meant to be filled with glue to fix the fibre in the connector.
This configuration avoid crimping the fibre since a sharp edge of the connector would easily damage it. The retaining clip (B) is screwed into
the wavelength shifting fibre connector (A) to allow the clear fibre connector (C) to be safely attached. All these pieces are non-standard and
a special mold was designed to produce them. 

\subsection{Photo-detectors}
The EMR has a dual readout. Each scintillator plane is equipped with a multi-anode PMT (MAPMT), measuring
the light output of individual bars, and a single-anode PMT (SAPMT), recording the integrated response of all bars in the plane. 

\begin{figure}[h]
 \centering
 \includegraphics[width=\textwidth]{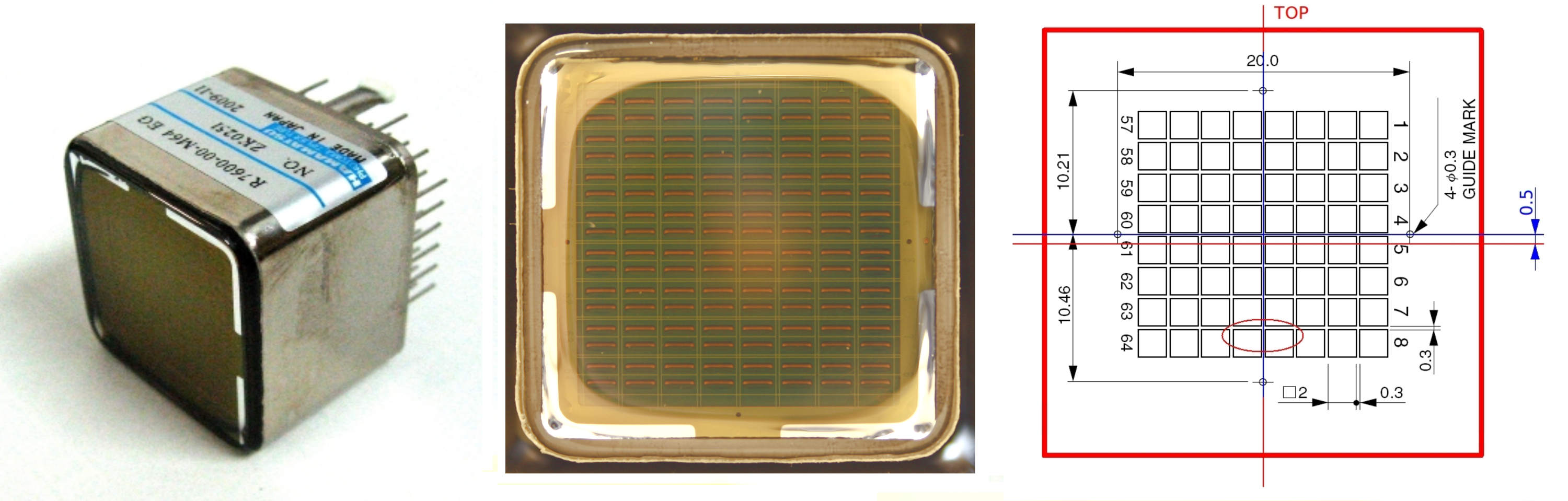}
 \caption[Multi-Anode PMT]{EMR Multi-Anode PMT: picture of the case and pins (left), anode matrix (centre) and anode matrix dimensions and offsets (right).}
 \label{fig:mapmt}
\end{figure}

\begin{figure}[h]
 \centering
 \includegraphics[width=0.45\textwidth]{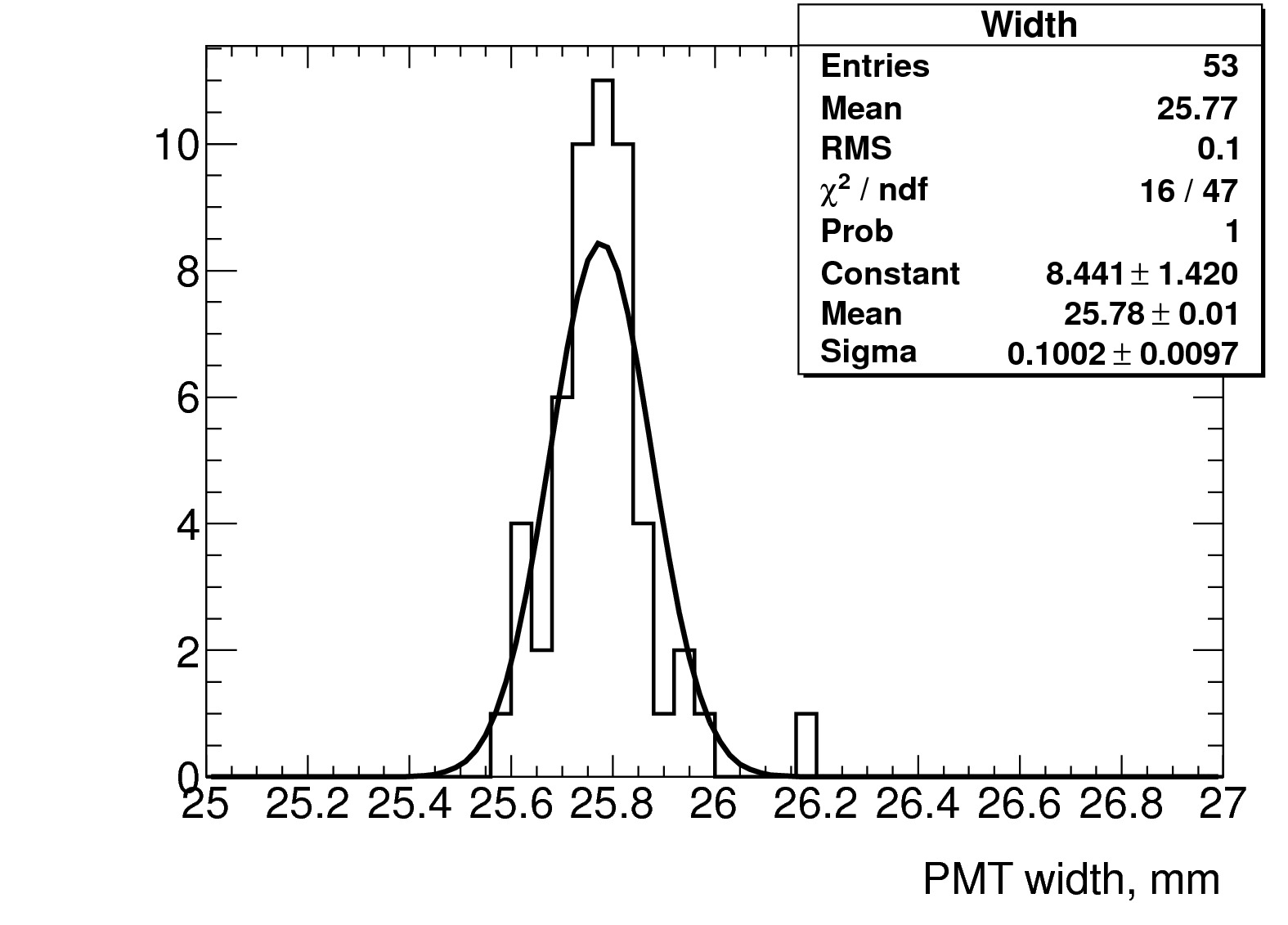}
 \includegraphics[width=0.45\textwidth]{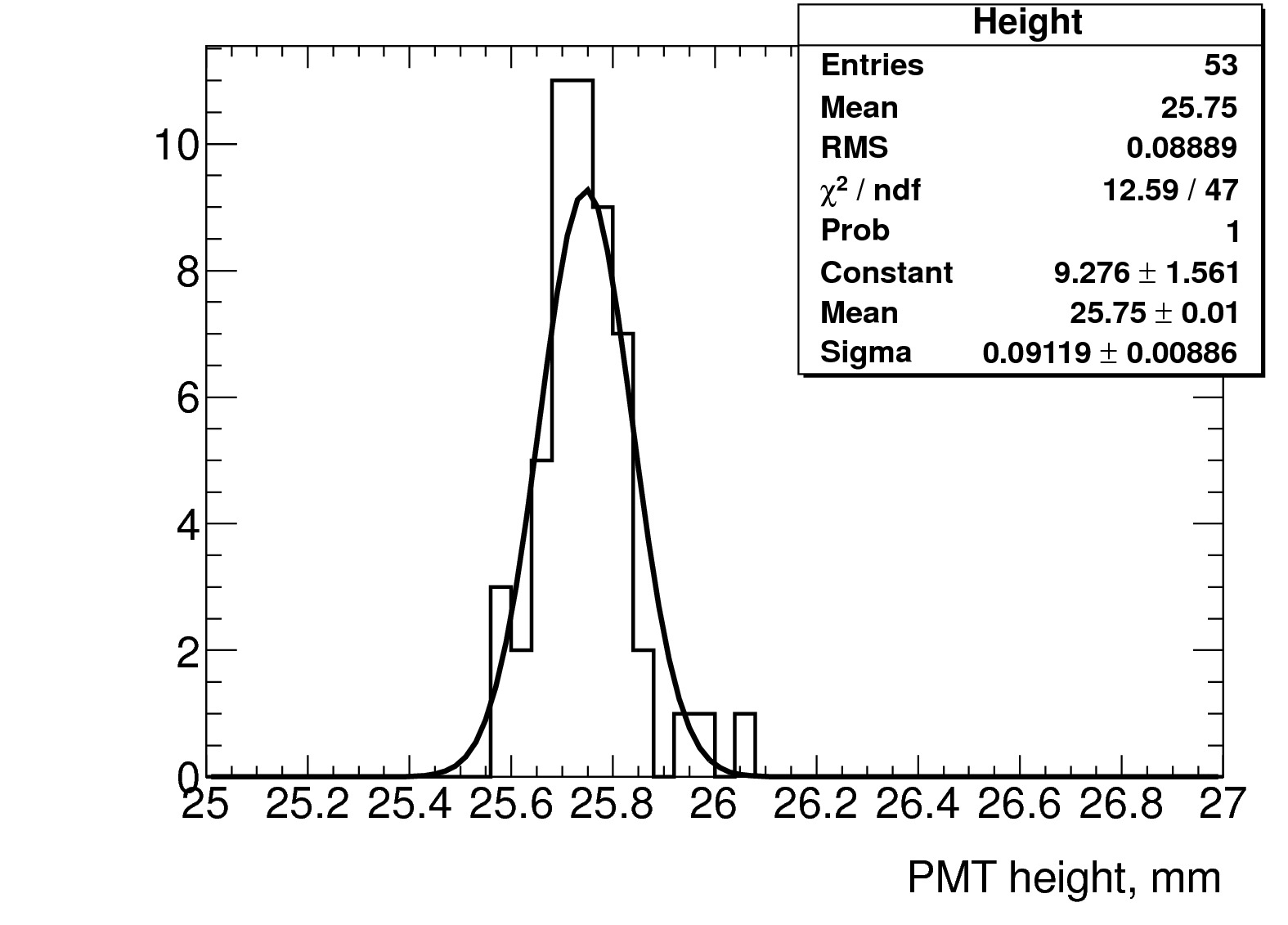}\\
 \includegraphics[width=0.45\textwidth]{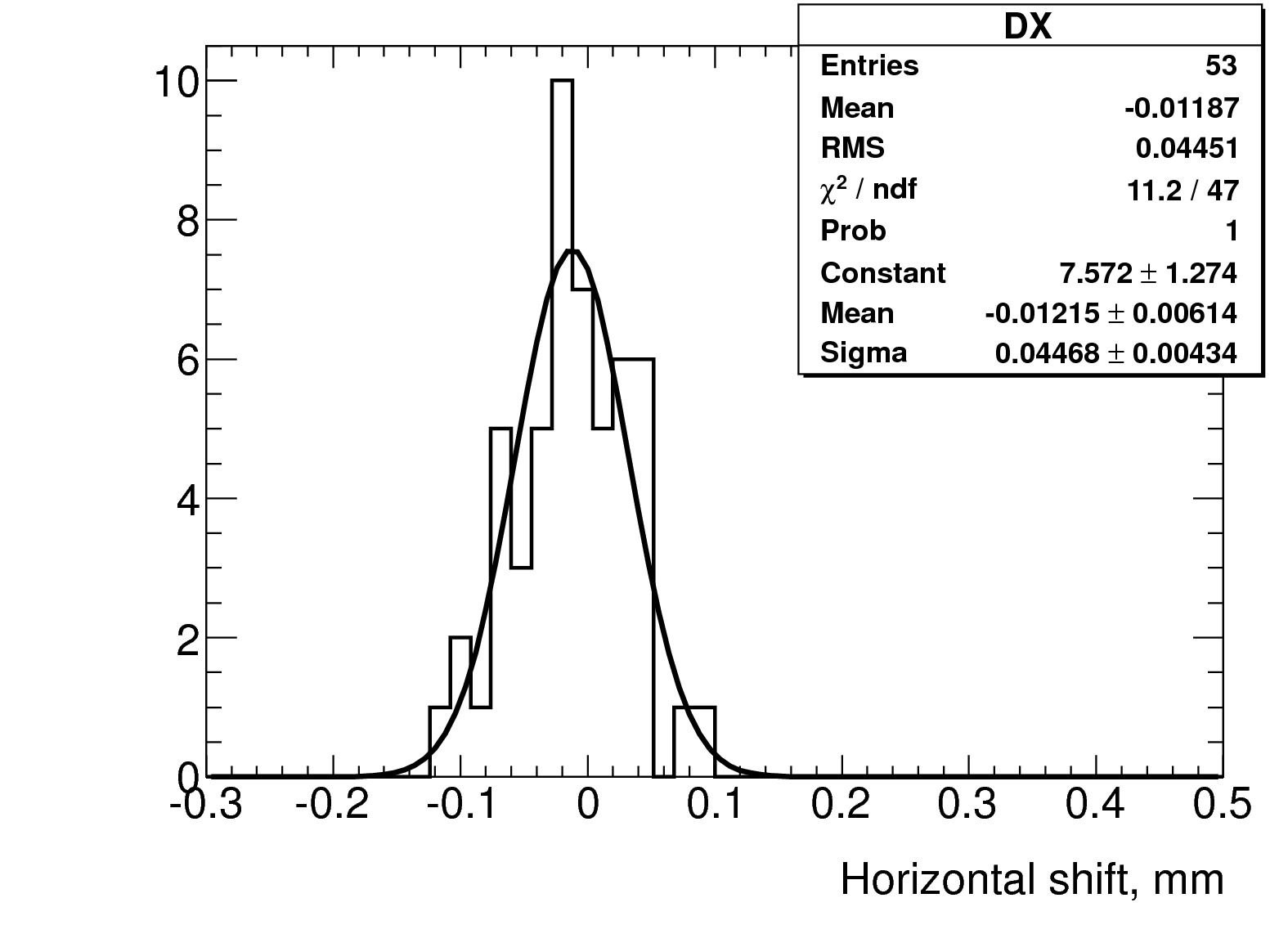}
 \includegraphics[width=0.45\textwidth]{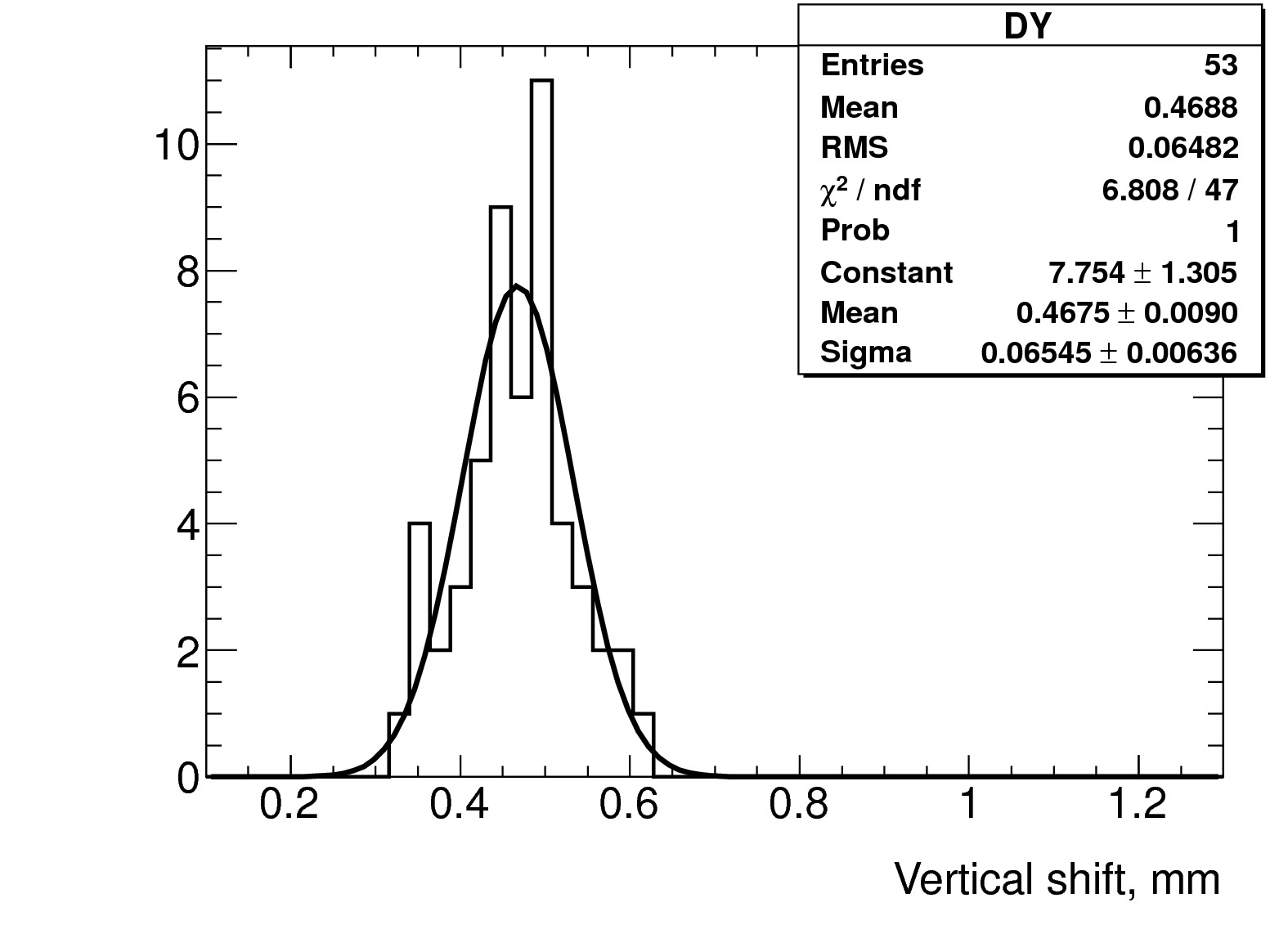}
 \caption[Distribution of Multi-Anode PMT dimensions]{Distribution of Multi-Anode PMT dimensions and offsets.}
 \label{fig:mapmt_dimensions}
\end{figure}

The MAPMT is a 64-channel PMT produced by Hamamatsu (model R5900-00-M64~\cite{hamamatsu_mapmt}, as pictured in figure~\ref{fig:mapmt}). The spectral
response (figure~\ref{qe}, bottom-left) matches the peak emission frequency of the wavelength shifting fibre. It is placed in a $\mu$-metal tube acting
as additional shielding against the fringe magnetic field. The PMT is aligned with respect to the fibre connector in such a way that each fibre shines
on one PMT channel. It is important to measure the dimensions of the PMT and particularly the position of the anode matrix with respect to the
PMT case. Figure~\ref{fig:mapmt_dimensions} shows the distributions of the measured dimensions (width and height) and displacements of the anode
matrix for 53 MAPMTs. On average, the matrix is shifted by 0.5\,mm upwards, which was taken into account in the design of the MAPMT fibre connectors.

The SAPMT is also placed in a $\mu$-metal casing. The EMR detector was initially assembled
using second hand SAPMTs, available after the disassembly of the HARP experiment \cite{harp}. They were 10-stage, linear-focused PMTs produced
by Philips (model XP2972). A special selection procedure was developed in order to select the best samples for the assembly of the detector
\cite{philips}. In 2014, during the upgrade of the detector, all Philips SAPMTs were replaced by new tubes produced by Hamamatsu (model R6427
\cite{hamamatsu_mapmt}, figure~\ref{qe} bottom-right).

\subsection{Electronics Layout}
In MICE the spill is defined as the short period following a target dip in the ISIS proton beam. The maximum spill rate allowed by the
MICE target system is $\sim\,$0.75\,Hz. The overall principle of the MICE Data Acquisition (DAQ) system is that, during the spill, the
accumulated digital data is kept in local memory buffers and the readout is performed once at the end of the spill.

\begin{figure}
 \centering
 \includegraphics[width=.99\textwidth]{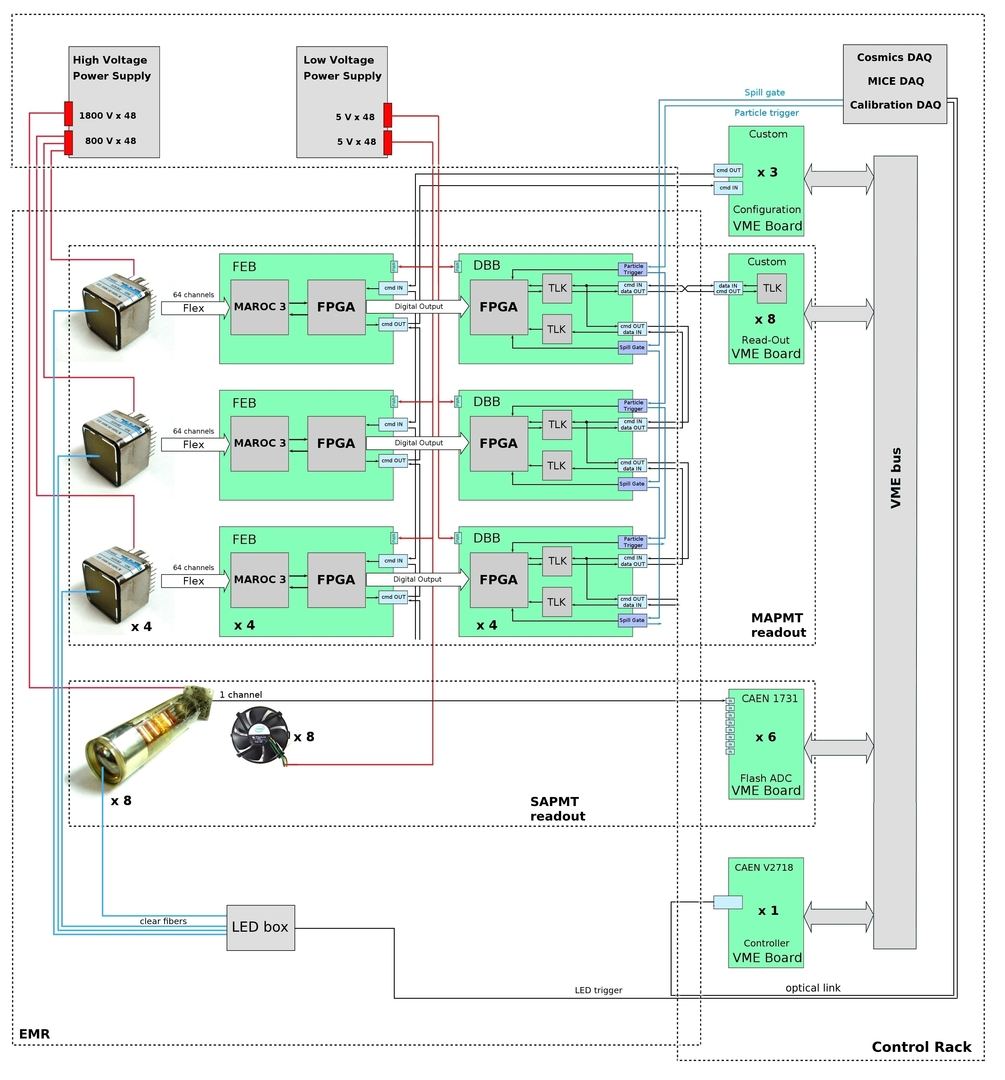}
 \caption[EMR electronics layout.]{EMR electronics layout. {\bf FEB:} front-end board for multi-anode PMT readout. {\bf DBB:} digitizer-buffer
 board. {\bf MAROC~3:} 64 channel readout ASIC for multi-anode PMT. {\bf MAPMT:} multi-anode PMT. {\bf SAPMT:} single-anode PMT.}
 \label{fig:EMR_electronics_layout}
\end{figure}

A schematic layout of the EMR electronics is shown in figure~\ref{fig:EMR_electronics_layout}. The multi-anode PMT is connected via a flex cable
to a front-end board (FEB), which processes the signals and sends them to a piggy-back digitizer-buffer board (DBB) for digitization and storage.
The FEB is configured by the VME configuration board (VCB), which resides in the VME crate in the control rack. Each VCB is able to configure
up to 16 FEBs, therefore three of them are required for the full detector. The DBBs are readout in groups of six. In each group the first DBB
is a master and the other five are slaves. All the six boards are daisy-chained via an ethernet cable and the master is connected to a VME readout
board (VRB), which transfers all the data from the six DBBs to the DAQ computer. In the whole detector there are 8 groups of DBBs, i.e. 8 VRBs are
installed in the control rack. 

\subsubsection{Front-End and Digitizer-Buffer Boards}

\begin{figure}[htp!]
 \centering
 \includegraphics[width=0.8\textwidth]{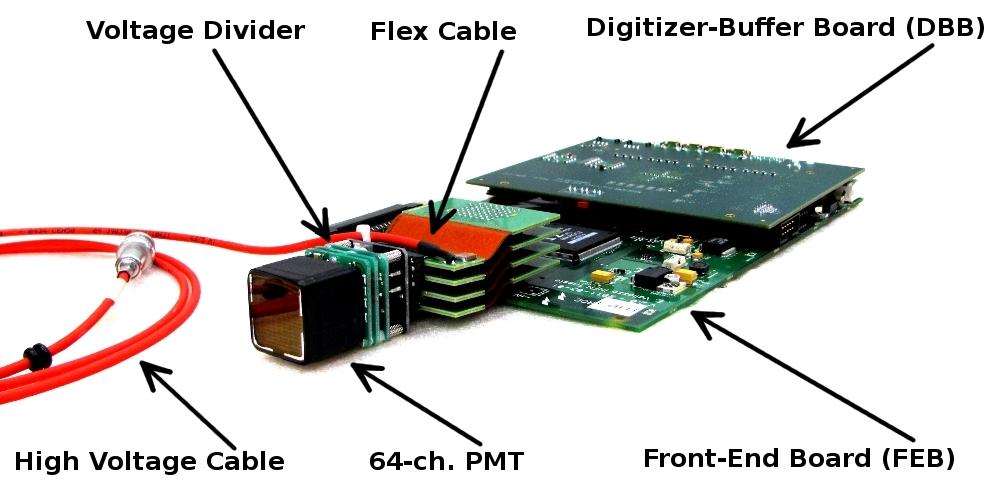}
 \caption[Front-end and buffer board assembly]{Front-end and Digitizer-Buffer boards assembly.}
 \label{fig:feb_dbb}
\end{figure}

The multi-anode PMT is readout by a dedicated front-end board equipped with a piggy-back digitizer-buffer board \cite{Bolognini2011108}, which stores
hit information during a spill. Figure~\ref{fig:feb_dbb} shows the full assembly that is mounted on each plane of the detector. It consists of a PMT
and its voltage divider connected to a FEB through a flex cable. The FEB processes all 64 the MAPMT signals using a 64-channel 
ASIC\footnote{Application-Specific Integrated Circuit} called MAROC\footnote{Multi Anode ReadOut Chip}\cite{maroc}.

The analogue signals are fed into the chip where they are processed in parallel. Each channel consists of a pre-amplifier with a variable gain,
a tunable slow shaper for analogue readout, a tunable fast shaper and a discriminator for the digital. The MAROC ASIC provides parallel digital outputs
forwarded to two high density connectors. The width of the discriminated signal represents the time-over-threshold measurement. One multiplexed
analogue output is also provided and this output is digitized by an external ADC\footnote{Analog Devices AD9220}. It takes  12.8 $\mu$s (64 channels
with a multiplexing clock of 5 MHz) to process the multiplexed signals, a time too long for the MICE DAQ duty cycle which foresees one particle every
5 $\mu$s during $\sim\,$1\,ms spills). Only the fast time-over-threshold measurement is used and recorded.

The function of the FPGA chip\footnote{Altera Cyclone II (EP2C35F484C8N)} is mainly to forward data from the MAROC to the DBB and to send configuration
signals from the VCB to the MAROC and verify their status. The board has a separate power supply for the analogue and the digital parts. The board is operated at
5V and has a power consumption of 3W.

The two essential roles of the DBB are to sample 63 of the channels coming from the FEB plus an external trigger signal and to store the accumulated
digital data during the spill. It also transmits these data upon request from the acquisition system. The digitization starts when the board
receives the "Spill Gate" signal from an external LEMO connector. The number of clock ticks from the beginning of the spill to the leading edge
and trailing edge of every discriminated signal coming from the FEB is recorded. The difference between two subsequent measurements represents
the time-over-threshold of the original signal. The clock sampling rate is 400\,MHz (2.5\,ns resolution). The external trigger signal is fed into
one specific input channel and is treated as any other signal. This signal does not serve as a trigger for the DBB itself, since the board records
continuously and all signals arriving within the spill gate are digitized and recorded. The timing of the trigger signal is important to identify
the hits that belong to a given particle and to match them with other detectors. The board also calculates the width of each spill, counts the number
of spills, the number of triggers in each spill and the number of hits in each channel.

The architecture of the DBB is organized around a single FPGA\footnote{Altera Stratix II (EP2S30F484C3N)} that performs the sampling, data
buffering and data-flow control functions of the board. The internal memory of the FPGA, configured as FIFO, is used to store the event data. Two
gigabit transceivers\footnote{TLK1501} are interfaced to the FPGA to provide the physical transmission channels and form an upstream command link and
a downstream data link. Six DBBs are grouped together and daisy-chained with upstream and downstream links via Ethernet cables. The first DBB in each
group is directly connected to the VRB via four coaxial cables.

\subsubsection{VME Configuration Board}
The VCB is a single FPGA\footnote{Altera Cyclone II (EP2C50F484C8N)} board designed to configure the MAROC chips on the FEBs. The communication between
the two boards is accomplished by LVDS\footnote{Low-Voltage Differential Signaling, communication protocol.} signals driven and received by LVDS
drivers/receivers connected to a corresponding FPGAs. The MAROC chip is configured by TTL\footnote{Transistor-Transistor Logic} signals composed of
830 bits coding the configuration parameters.The VCB would in principle allow the readout of the MAROC boards analog output but this is not implemented
in the current design. The board communicates with the DAQ computer through the VME bus via the VME controller. 

\subsubsection{VME Read-Out Board}\label{electronics:subsec:vme_readout_board}

One VRB performs the readout of a group of six DBBs. It is a single FPGA\footnote{Altera Cyclone II (EP2C50F484C8N)} board with a gigabit 
transceiver\footnote{TLK1501}, which drives the communication with the DBBs and four high-speed 16M-bit static RAMs\footnote{IS61WV102416BLL-10TLI
- SRAM, 16Mbit, 10ns, 48TSOP}, providing a local memory buffer. The total size of the digital data, accumulated in six DBBs during the spill can vary
significantly, depending on the configuration of the beamline channel. Typically this size does not exceed 25 kB, including noise records. During the
readout cycle, the data transfer between the DBBs and the DAQ computer is executed in two steps. After a request from the DAQ computer, the VRB starts
transferring data from the 6 DBBs. The gigabit transceiver is used for this and the received data is temporarily stored locally. The four static RAMs,
organized as 16 bits data words, are grouped in two pairs, providing the record of the DBB data, originally structured in 32 bits data words. Once the
first part of the transfer is completed and all the data accumulated by the 6 DBBs during the spill are available in the local memory buffer of the VRB,
the DAQ computer sends a second request which triggers the transfer of these data over the VME bus.

\subsubsection{Flash ADC Board}\label{electronics:subsec:fast_adc_board}

A waveform digitizer V1731, made by CAEN~\cite{V1731} is used to read out signals from the single-anode PMTs. The digitizer has a sampling frequency
of 500\,MHz (2\,ns timing resolution). The pulse shape of the input signal is digitized by an 8-bit ADC and continuously written in a circular memory
buffer. When a trigger is received, the FPGA writes a certain number of samples into the buffer, which then is available for readout via the VME bus.

\subsection{Mechanics}\label{design:subsec:mechanics}

\begin{figure}[htp!]
 \centering 
 \includegraphics[width=\textwidth]{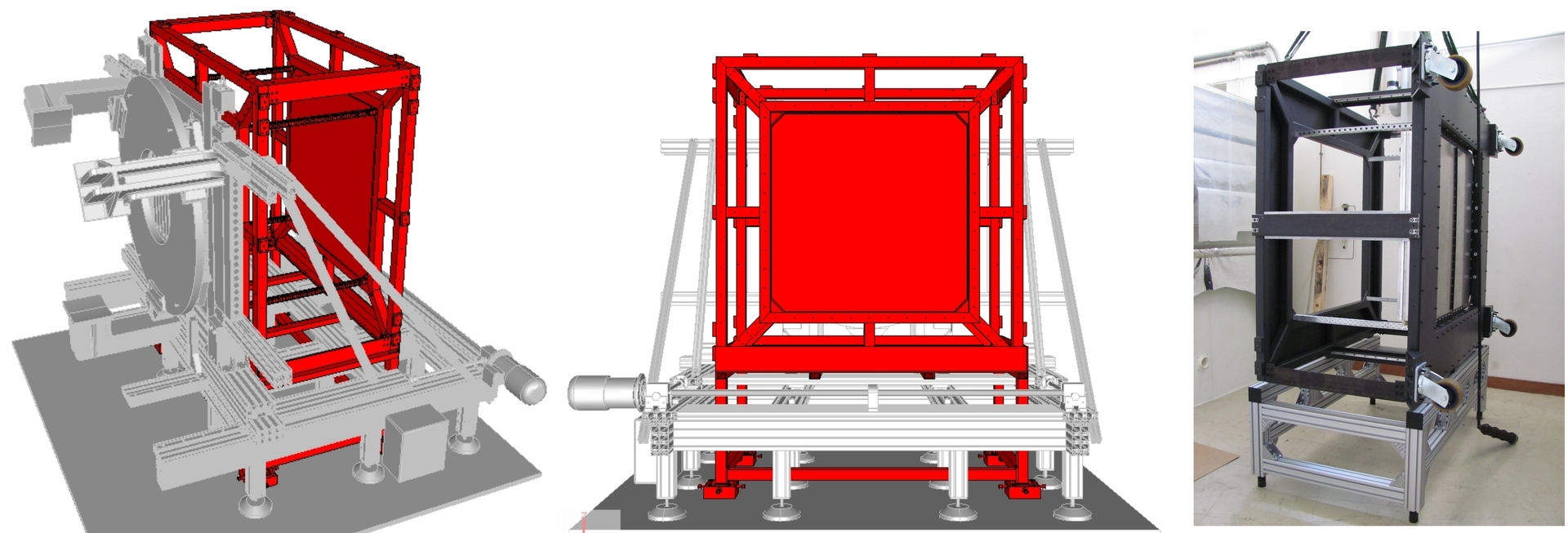}
 \caption[EMR support frame]{EMR support frame. When installed in the MICE experimental hall, the EMR is integrated into the support structure of the other 
 downstream particle identification detectors.}
 \label{fig:emr_support}
\end{figure}

The total weight of the sensitive volume of the detector is almost 1\,tonne. During construction and installation, it was required to be rotated and
transported from Geneva to the UK. A reinforced support frame, represented in figure~\ref{fig:emr_support}, was designed to withstand the weight of the
sensitive detector and the stress coming from the transportation and installation. In its final position, the EMR is installed such that planes are
located perpendicular to the beam direction. 

\begin{figure}[htp!]
 \centering
 \includegraphics[width=0.9\textwidth]{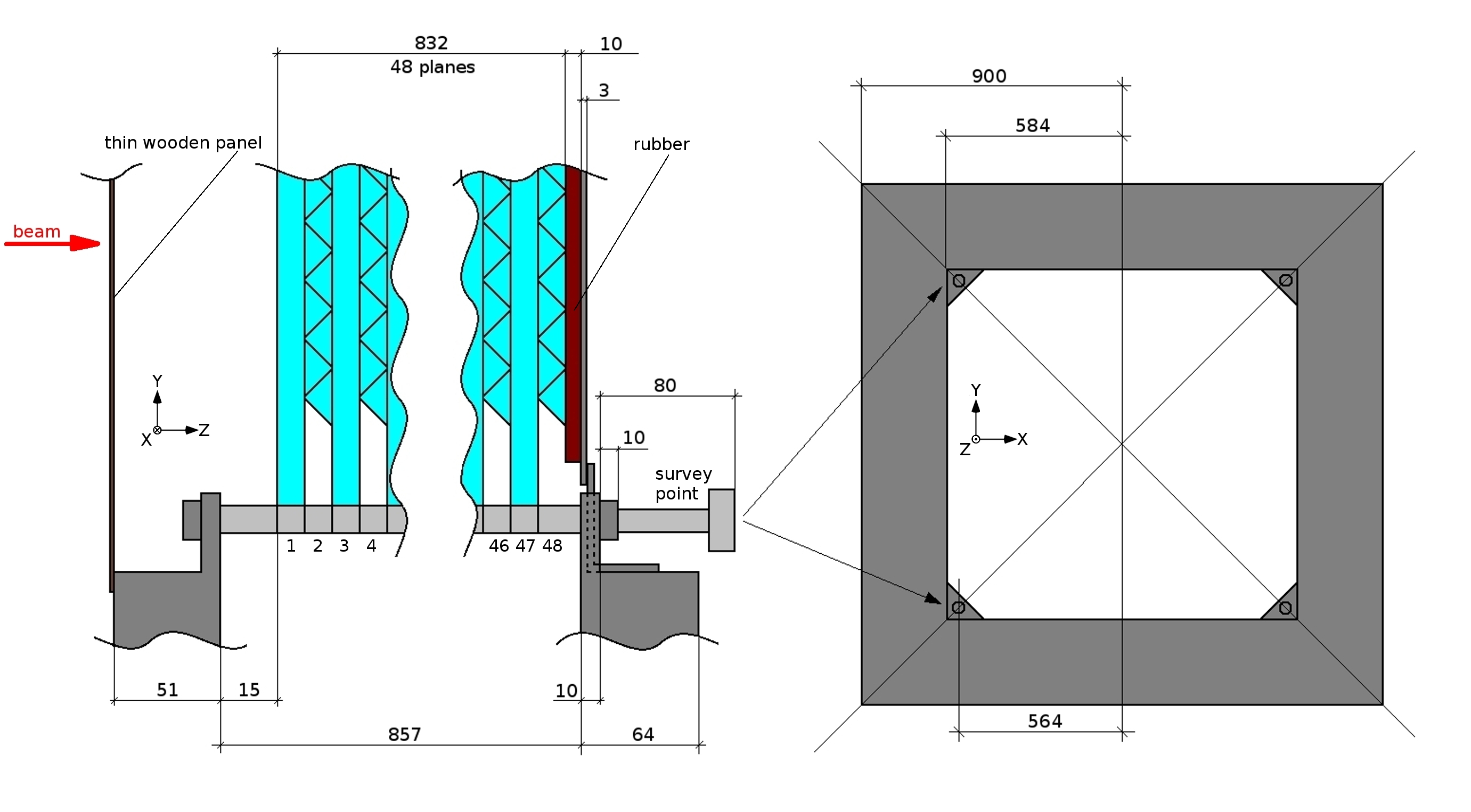}
 \caption[EMR dimensions]{EMR dimensions [mm].}
 \label{fig:internal_dimensions}
\end{figure}

Figure~\ref{fig:internal_dimensions} shows the location of the sensitive volume with respect to the support frame. The frame is covered with panels so that
the entire detector is light-tight. A 5\,cm-thick iron plate is used as magnetic shielding (total weight of $\sim\,$755\,kg). The opening in the shielding
panel, which matches the size of the sensitive volume, is closed with a thin rubber end cap. The back panel is closed with a metal end cap.

\section{Construction}\label{sec:construction}

All the construction work was done at the University of Geneva. As exposure to ultra-violet (UV) light or high temperature can damage the polystyrene
molecules, activities related to fibres and scintillators were performed in a UV clean room, i.e. lights and windows were covered with UV-protective
films, while air conditioning kept the temperature around 25\,\degree C. 

The first step in the construction was to glue the wavelength shifting fibres
into the bars. Transparent epoxy\footnote{Prochima E30 water effect resin} was used to glue the WLS fibre in order to increase the light collection
efficiency. Although 2832 bars were required for building the full detector, 3150 bars were glued and assembled with fibre connectors in order to
provide enough spares. Both faces of the bar's fibre connectors were polished with a custom polishing machine. Four different grades of sand paper
were used to achieve a mirror like quality of the polished surfaces. The last step was performed using a 1\,$\mu$m grade diamond-based polishing paper.

\begin{figure}[htp!]
 \centering
 \includegraphics[width=\textwidth]{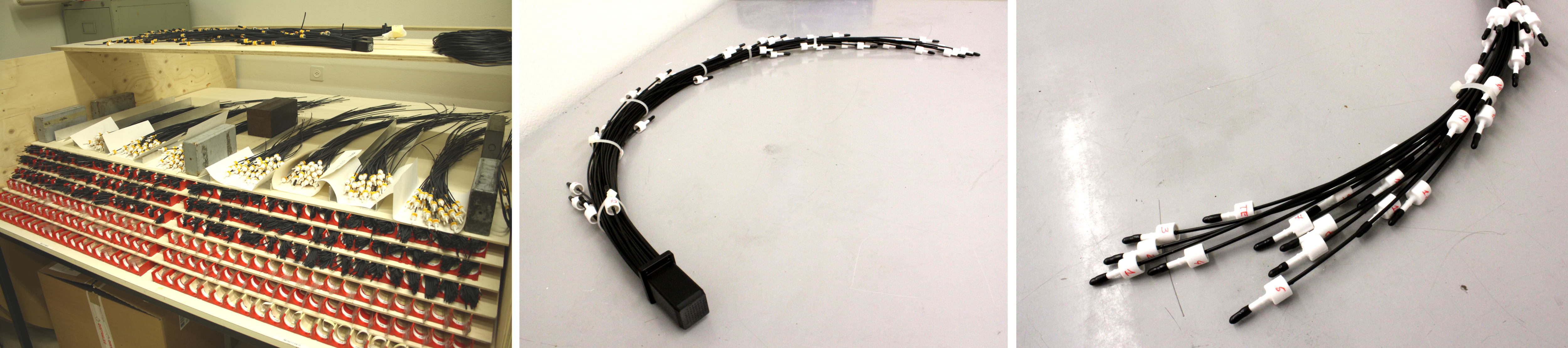}
 \caption[Clear fibre bundles]{Clear fibre bundle construction: fibres cut to the right lengths (left), an MAPMT connector (centre) and its fibre
 connectors (right).}
 \label{fig:fibre_bundles}
\end{figure}

Fiber bundles made of 60 clear fibres (see figure~\ref{fig:fibre_bundles}) were manufactured. In the bundle, each fibre has an individual length, providing
a maximum bending radius when connected. A fibre connector (see figure~\ref{fig:fibre_connectors_cad}, bottom right) is glued at one end of each fibre. At
the other end all fibres were glued either in multi-anode or single-anode PMT connectors (see figure~\ref{fig:pmt_connectors}). Once glued, both fibres and
connectors were polished on a bench, similar to the one used to polish all bar connectors. In total 96 fibre bundles were produced (48 per type of PMT).

\subsection{Quality Assurance Tests}\label{construction:subsec:quality_tests}

Several quality tests were implemented, in order to assure the best possible performance of the different components of the detector. 

A dedicated bar quality test bench was built in order to test the light transmission of each bar, including the transmission of the WLS fibre and the
quality of the two connectors. The test bench consisted of a LED\footnote{Light Emitting Diode} system, a holder for 4 scintillator bars and a digital
camera placed in a light-tight box. The LED system included a blue LED source, a light mixing box and diffusers to provide a homogeneous light signal
in the four bars. The camera takes a photo of the four connectors at the opposite side of the bars. One of the bars is considered to be a reference to
which the other measurements are normalized to. This takes into account the effect of any LED instability and allows to compare different measurements. 

\begin{figure}[htb]
 \centering
 \includegraphics[width=.8\textwidth]{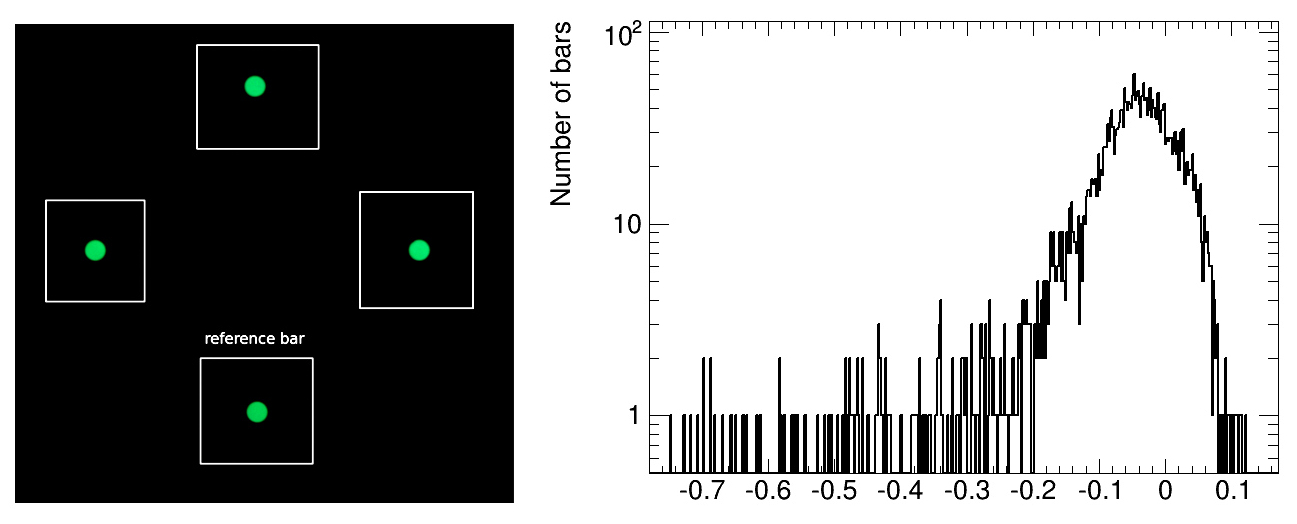}
 \caption[Bar quality test measurements]{Bar light transmission test: single measurement of four bars (left) and the distribution of the relative residual
 luminosity (right).}
 \label{fig:measurements}
\end{figure}

An automated program was used to analyse the photos (figure~\ref{fig:measurements}, left) and calculate the luminosity of each bar. The right plot in
figure~\ref{fig:measurements} shows the distribution of the measured relative residuals of the light output with respect to the reference bar. The relative
residual is defined as the difference between the measured value and the average value normalised by the average. Only bars with a relative residual
intensity above -0.15 were accepted for the plane assembly. A fraction of 9.7\% of the 3150 bars did not pass that requirement and were rejected.

\begin{figure}[htb]
 \centering
 \includegraphics[width=0.8\textwidth]{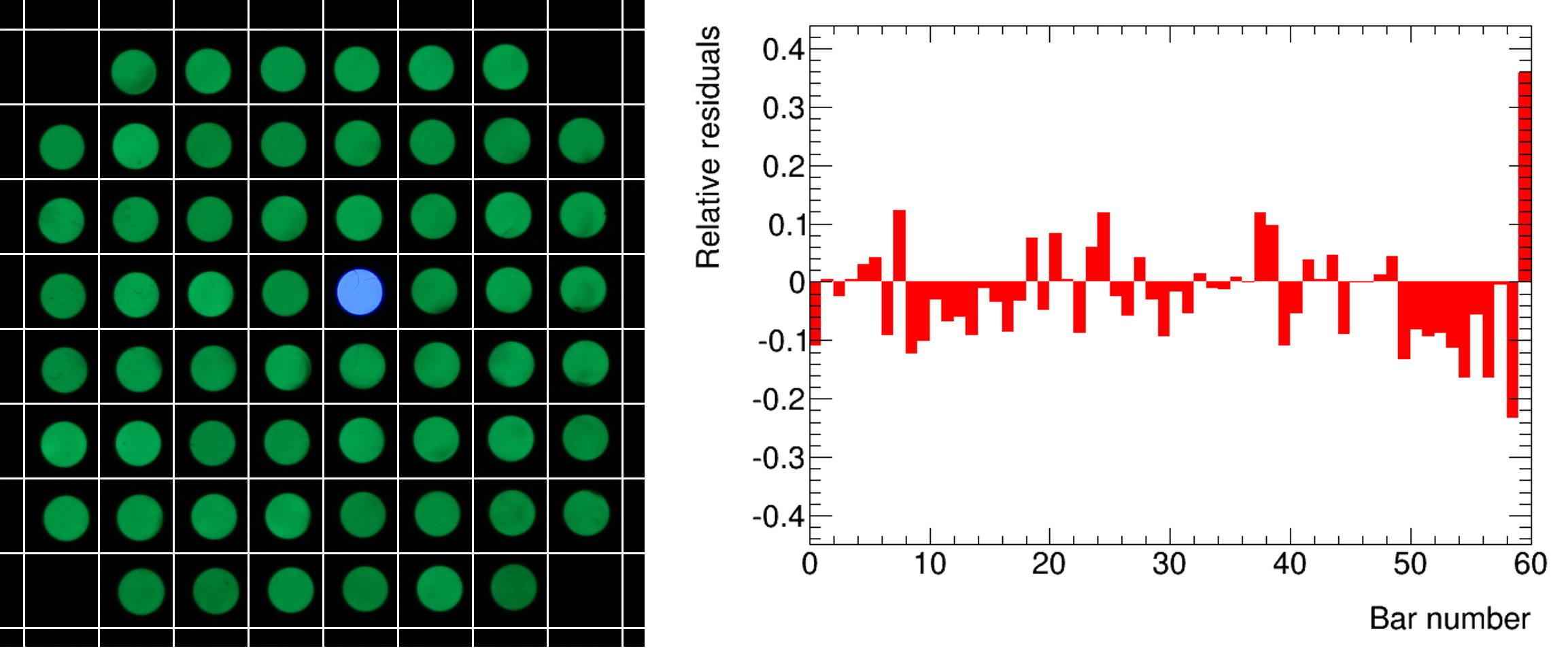}
 \caption[Example of plane quality tests]{Example of a single plane quality test: picture of the PMT fibre mask (left) and relative residual luminosity
 of the 60 fibre outputs (right). The 60th channel is the test channel.}
 \label{fig:plane_tests_one_planes}
\end{figure}

A similar test was used to examine each EMR plane after the assembly~\cite{emr_elquality}. A LED tube attached to a single-anode PMT connector was used
to send light through the fibres (WLS and clear) to the multi-anode PMT connector, where a picture of the PMT mask was taken by a camera. An automated
program was used to analyse the photos (figure~\ref{fig:plane_tests_one_planes}, left) and calculate the relative residuals of the light intensity
of the individual channels as shown for a single plane in the right plot of figure~\ref{fig:plane_tests_one_planes}. This test verified the light
transmission of the fibre bundles, but also the quality of the interconnections between the WLS fibres and clear fibres. A plane was accepted only if
the relative residual intensities of all 60 channels were above -0.4.

\begin{figure}[htb]
 \centering
 \includegraphics[width=0.45\textwidth]{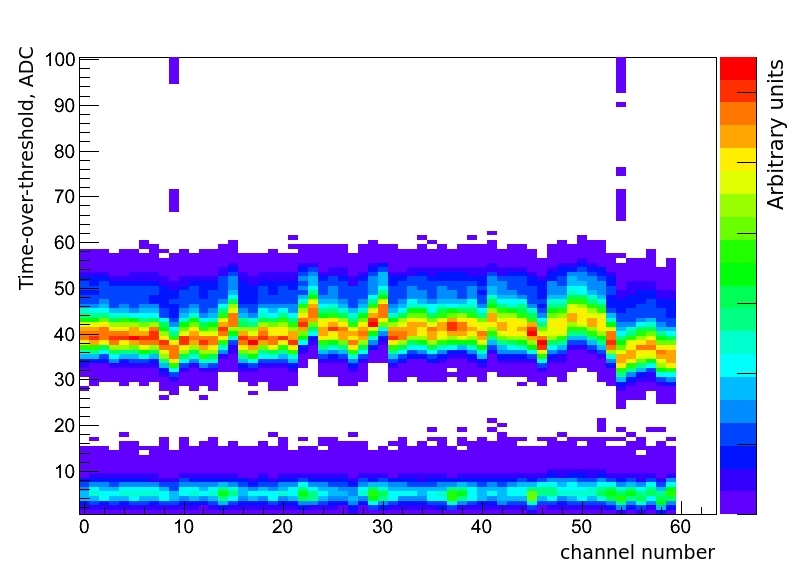}
 \includegraphics[width=0.45\textwidth]{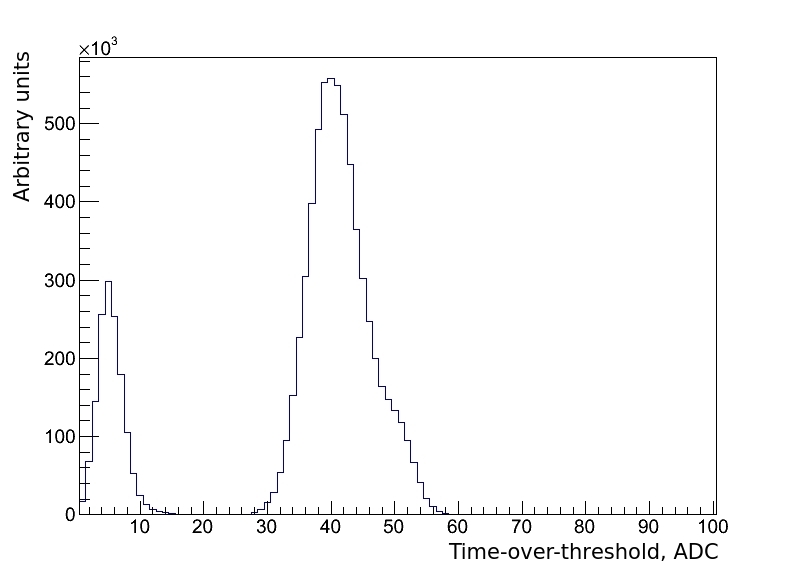}\\
 \includegraphics[width=0.45\textwidth]{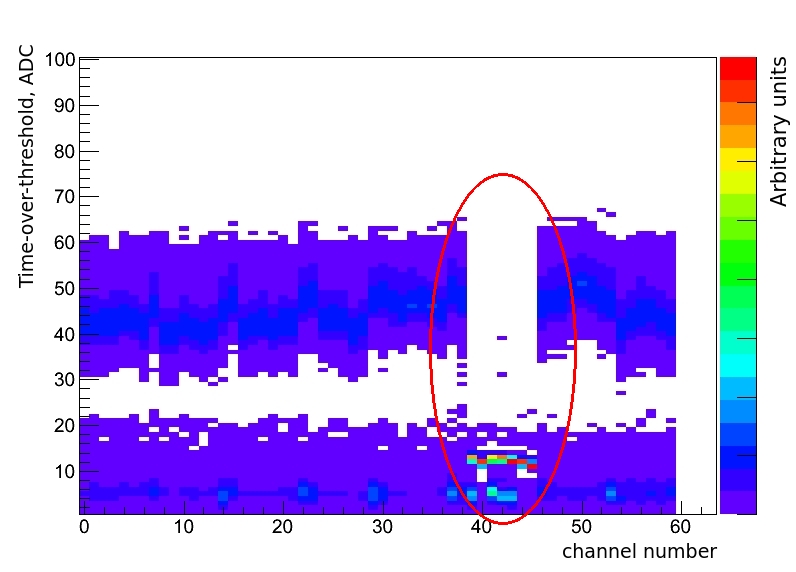}
 \includegraphics[width=0.45\textwidth]{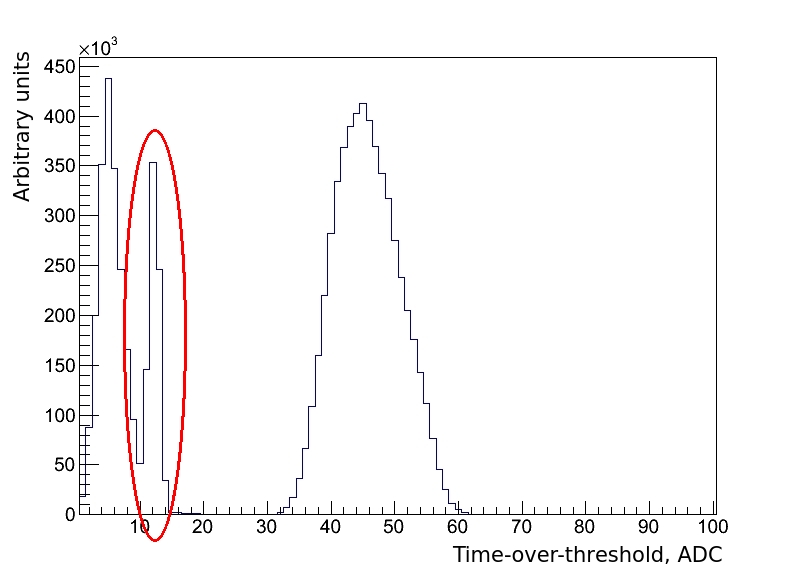}
 \caption[Electronics quality tests]{MAPMT readout quality tests of a functional board (top) and a faulty board (bottom). The left plots show the 
 time-over-threshold distribution as a function of the channel number and the right-hand plots represent the integrated distribution of all channels
 in a given board.}
 \label{fig:tot_feb_dbb_test}
\end{figure}

A separate test bench was set up in order to verify the functionality of the three major components of the EMR electronics: the multi-anode PMTs, the 
front-end boards and the digitizer-buffer boards. It reproduced the full electronics chain used to readout the detector with the only difference that the
light was generated by a LED source, powered by a variable amplitude pulser. The LED was attached to the MAPMT injecting light in all channels at the
same time. The final measurement that is provided by the system is a time-over-threshold (ToT) of the PMT signal. During the tests this measurement was
used as a figure of merit to characterize the electronics chain (PMT, FEB and DBB). The top of figure~\ref{fig:tot_feb_dbb_test} shows an example of a
fully functional electronics chain and the bottom an example of a faulty board. Boards exhibiting the former behaviour were accepted for installation in
the detector.

\subsection{Examining the performance with the LED calibration system}\label{sec:led_perf}
After the assembly was completed, the detector was fully powered and tested during a few days of LED and cosmics data collection. As described above,
the detector is equipped with a built-in LED system for calibration.  Light is transported from a LED driver through 96 clear fibres to each SAPMT and to
one specific channel of each MAPMT (test channel). 

\begin{figure}[htr!]
  \begin{minipage}[b]{.48\textwidth}
   \centering
   \includegraphics[width=\textwidth]{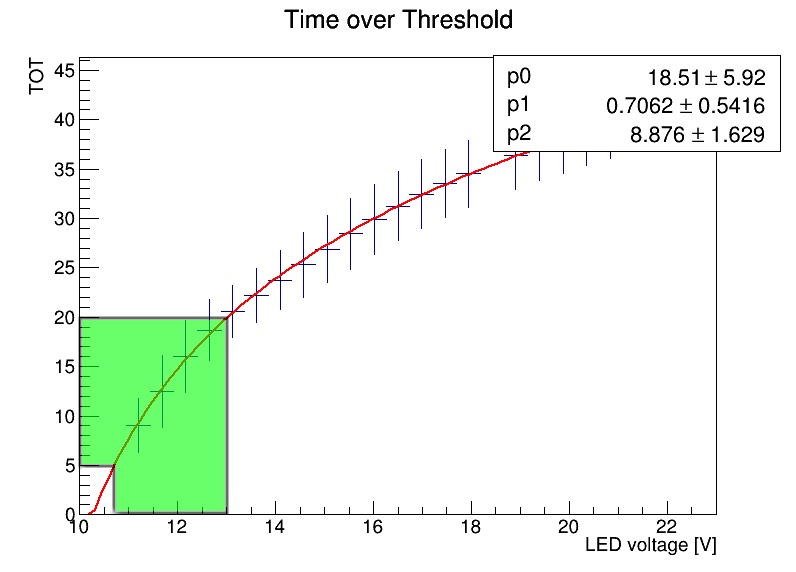}
   \caption{Mean time-over-threshold (ToT) in the test channel as a function of the LED driver voltage. The fit shows a logarithmic relation of the form
   ToT$=p_0\ln\left[p_1V+p_2\right]$.}
   \label{fig:voltage}
  \end{minipage}
  \hfill
  \begin{minipage}[b]{.48\textwidth}
   \centering
   \includegraphics[width=\textwidth]{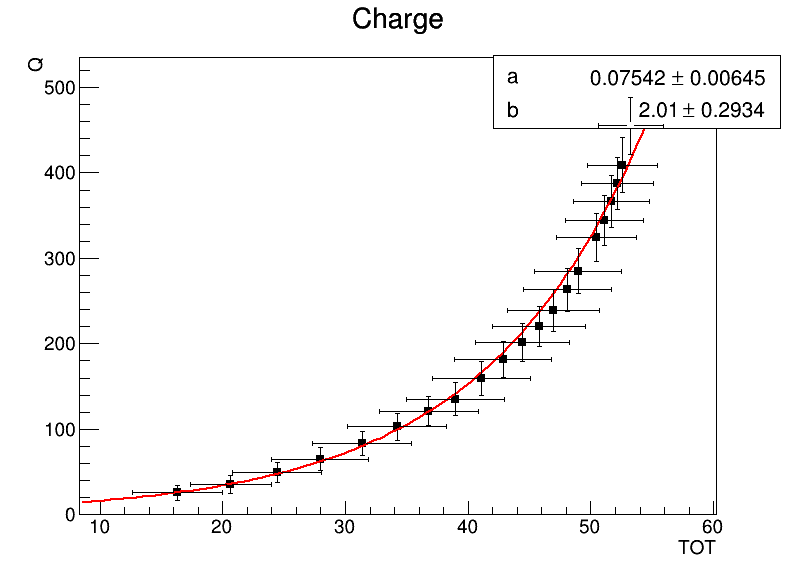}
   \caption{Plane charge as a function of time-over-threshold for different settings. The error bars represent the distributions RMS. The fit shows a
   relation of the form $Q=\exp(a$ToT$+b)$.}
   \label{fig:charge_tot}
   \end{minipage}
\end{figure}

The LED driver was tuned with a variety of voltages ranging from 11.0 V to 22.0 V in steps of 0.5V. For each setting, 10000 pulses were recorded. The
mean time-over-threshold in the test channel of a MAPMT is represented as a function of the LED driver voltage in figure~\ref{fig:voltage}. The green
area represents the voltage region for which the recorded ToT is consistent with the signature of a minimum ionizing particle.

The collected data were used to investigate the dependence of the ToT measurement on the original total charge of the MAPMT signal. This can be done,
assuming that the total charge of the signal in the test channel of the MAPMT is proportional to the total charge $Q$, recorded by the SAPMT when both
receive light from the same LED pulse. An exponential behaviour of the form
\begin{equation}
 Q=\exp\left[a\times ToT+b\right]
 \label{exp}
\end{equation}
is expected, with $a$ and $b$ being two unknown parameters. The parameter $a$ gives the slope of the exponential in log scale and depends on the EMR 
characteristics such as the scintillation time constant, the FEB shaping function or the threshold level. As a result, we expect this parameter to be
constant for each plane with small variations. The parameter $b$, on the other hand, depends on the two PMTs gain and can vary significantly from one
plane to another. These two parameters were obtained experimentally by fitting the relation $Q$ vs. ToT for each individual plane of the detector.
Figure~\ref{fig:charge_tot} shows the exponential relation between time-over-threshold and charge in one of the EMR planes.

\subsection{Examining the performance with cosmic rays}\label{sec:cosmic_perf}
Cosmic rays present an ideal source of particles that can be used to characterize, debug and tune the detector. Cosmic rays that reach the detector are
typically multi-GeV muons that cross EMR without stopping. An externally generated 3\,ms signal was used in order to reproduce the timing structure
of the "Spill" gate signal, used by the MICE DAQ system. A coincidence between the SAPMT signals from two planes and the Spill gate signal was used
as a trigger.

\begin{figure}[!htr]
\centering
\includegraphics[width=\textwidth]{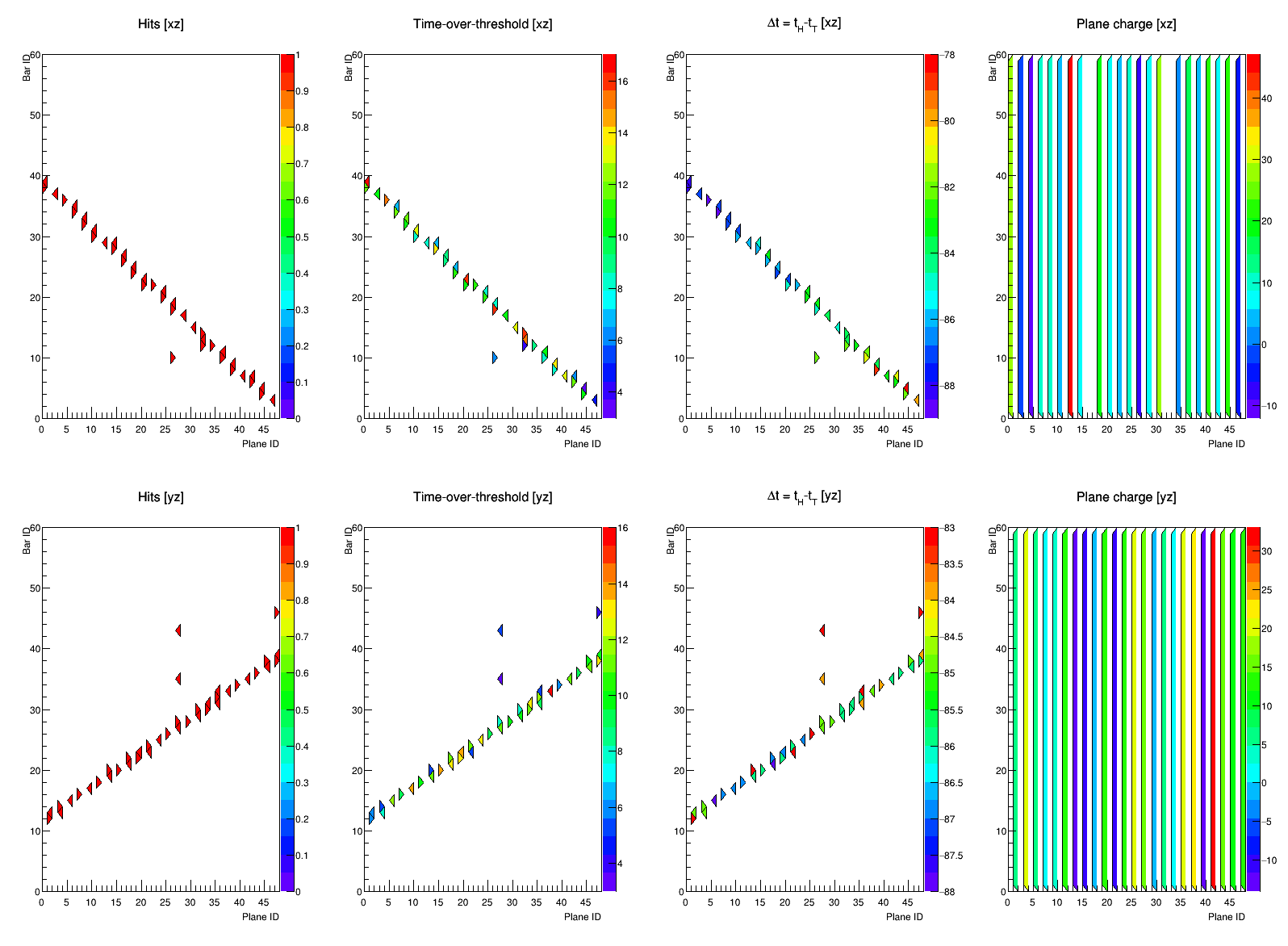}
\caption{Cosmic muon event in the EMR in the $xz$ (top) and $yz$ (bottom) projections. From left to right: hits per bar, ToT, hit time
with respect to the trigger and plane charge.}
\label{fig:cosmic_muon}
\end{figure}

The EMR planes were perpendicular to the ground at the time of this test. Data were taken for 60 hours and yielded $\sim2.23\times10^5$ triggers. The raw
measurements of the EMR comprise the hit time and ToT for each bar and the integrated plane charge. The two planes used as trigger did not record a plane
charge. The number of hits recorded in each bar was of the order of $\sim10^3$. The test revealed only 5 dead channels ($\sim$0.2\,\% of the whole detector).
A typical cosmic event is shown in figure~\ref{fig:cosmic_muon}.

\subsubsection{Cosmic muon signature}\label{sec:cosmic_sig}
Cosmic muons are minimum ionizing particles and hence deposit on average the same amount of energy per unit length. The muon typically hits two triangular
bars in a plane, by design. The additional hits come either from crosstalk or noise (e.g. third hit in plane 26 in figure~\ref{fig:cosmic_muon}, top-left).
All hits from cosmic muons have approximatively the same time offset with respect to the trigger. To separate the true hits from the noise, a cut on
the time difference between a bar hit and the trigger was applied. A thorough study of the crosstalk was performed and is presented in section~\ref{sec_xt}.
 
\begin{figure}
 \centering
 \begin{minipage}[b]{.48\textwidth}
  \centering
  \includegraphics[width=.9\textwidth]{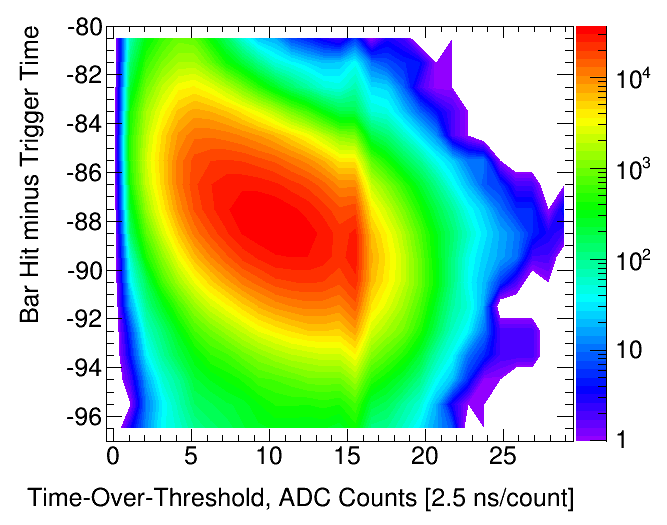}
  \caption{Energy and time structure of the hits recorded in the MAPMT for cosmic muons (MIP signals).}
  \label{fig:tot_vs_deltat}
 \end{minipage}
 \hfill
 \begin{minipage}[b]{.48\textwidth}
  \centering
  \includegraphics[width=\textwidth]{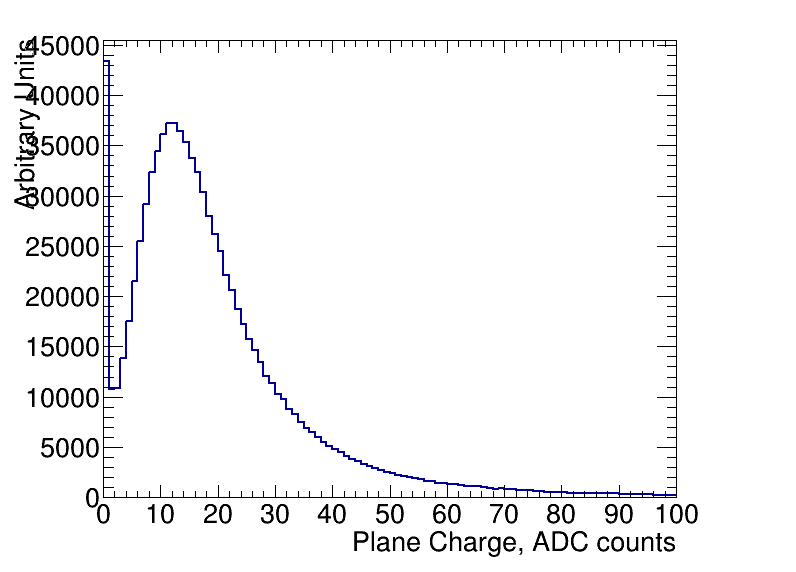}
  \vspace{1mm}
  \caption{Distribution of the charge recorded in the SAPMT for cosmic muons (MIP signals).}
  \label{fig:plane_charge}
 \end{minipage}
\end{figure}

Figure~\ref{fig:tot_vs_deltat} shows the time and ToT structure of the hits recorded in the MAPMT for the whole sample of cosmic data. The hits come prior
to the trigger in the readout chain due to the delay in the logic box. Time walk affects the timing so that the lower amplitude signals come later on
average. The distinctive spike present at $\sim15$\,ADC counts in the ToT distribution is a feature of the cosmic data acquired with detector planes
perpendicular to the ground and the non-linear connection between charge and ToT.

The SAPMTs were used to measure the total charge in planes. The digitizers store waveforms of signals associated to triggers within a certain acquisition
window. These waveforms are integrated in a signal window off-line taking into account the pedestal positions in order to compute the total plane
charge, as shown for the whole cosmic data sample in figure~\ref{fig:plane_charge}.

\subsection{LED-based crosstalk analysis}\label{sec_xt}
The EMR is susceptible to two types of crosstalk: optical crosstalk, i.e. a single fibre of a bundle shining on more than one channel of the MAPMT mask,
and anode crosstalk, i.e. a photo-electron leaking from a dynode to an adjacent accelerating structure. An analysis was developed in \cite{emr_xt, Francois}
to evaluate the significance of this phenomenon.

Cosmic or beam data are poorly suited for this analysis. A real particle often hits two bars or more within each plane, which makes it impossible to
disentangle real signals from crosstalk in neighbouring channels. A LED calibration system is a more reliable tool to drive the analysis. The test channel,
connected to the LED light source, has four directly adjacent channels: top (N), bottom (S), left (W) and right(E). This method ensures that hits in the
adjacent channels are caused by crosstalk only.

The first parameter that characterizes the crosstalk is the charge ratio $R_Q^i$, between the signal amplitude in an adjacent channel $i$ and the primary
amplitude in the test channel. As explained already, the charge is not measured directly but is related to the ToT through equation~\ref{exp}. For the ratio
we have:
\begin{equation}
R_Q^i=\frac{Q_i}{Q_0}=\frac{\exp[aToT_i+b]}{\exp[aToT_0+b]}=\exp\left[a(ToT_i-ToT_0)\right].
\end{equation}

\begin{figure}[htr!]
  \begin{minipage}[b]{.45\textwidth}
   \centering
   \includegraphics[width=\textwidth]{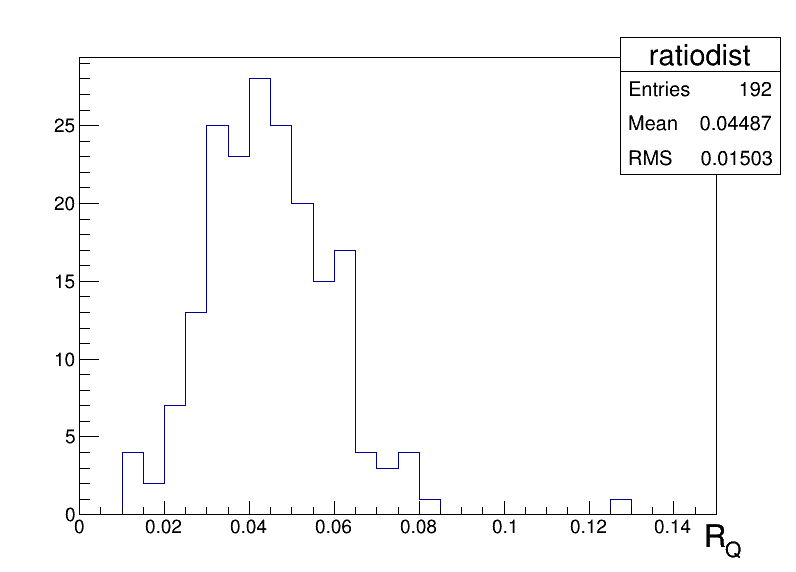}
   \caption{Fraction of the original charge that can leak in adjacent channels.}
   \label{fig:ratio_dist}
  \end{minipage}
  \hspace{3mm}
  \begin{minipage}[b]{.45\textwidth}
   \centering
   \includegraphics[width=\textwidth]{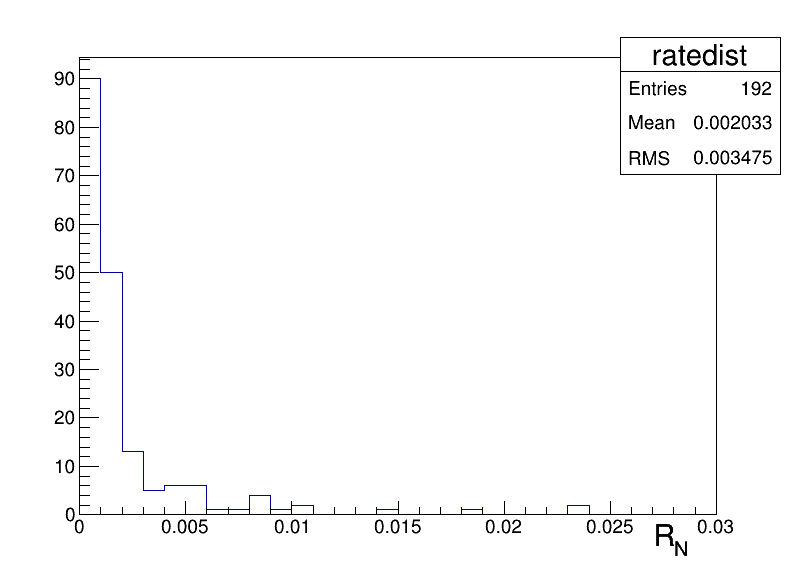}
   \caption{Fraction of the time a signal produces crosstalk for a typical MIP energy loss.}
   \label{fig:rate_dist}
  \end{minipage}
\end{figure}

$R_Q^i$ is measured for the maximum LED voltage setting as the resolution evolves as $1/\sqrt{Q}$. The ratio measured in the 192 readout channels
(directly adjacent N,S,W,E channels of each plane) is represented in figure~\ref{fig:ratio_dist}. The fraction of the original signal that typically leaks in
adjacent channels is $4.49\pm0.11\,\%$.

The second parameter used to characterize the crosstalk is the rate. The measured quantity is the ratio $R_N^i$ of hits in a surrounding channel $i$ to the
total amount of pulses generated in the test channel. This quantity is measured in the 192 readout channels for a voltage setting in the green area of figure
\ref{fig:voltage}, corresponding to MIP-like signals, and represented in figure~\ref{fig:rate_dist}. The average rate fraction is $0.20\pm0.01\,\%$, well
within design requirements.

\begin{figure}[htr!]
   \centering
   \includegraphics[width=.4\textwidth]{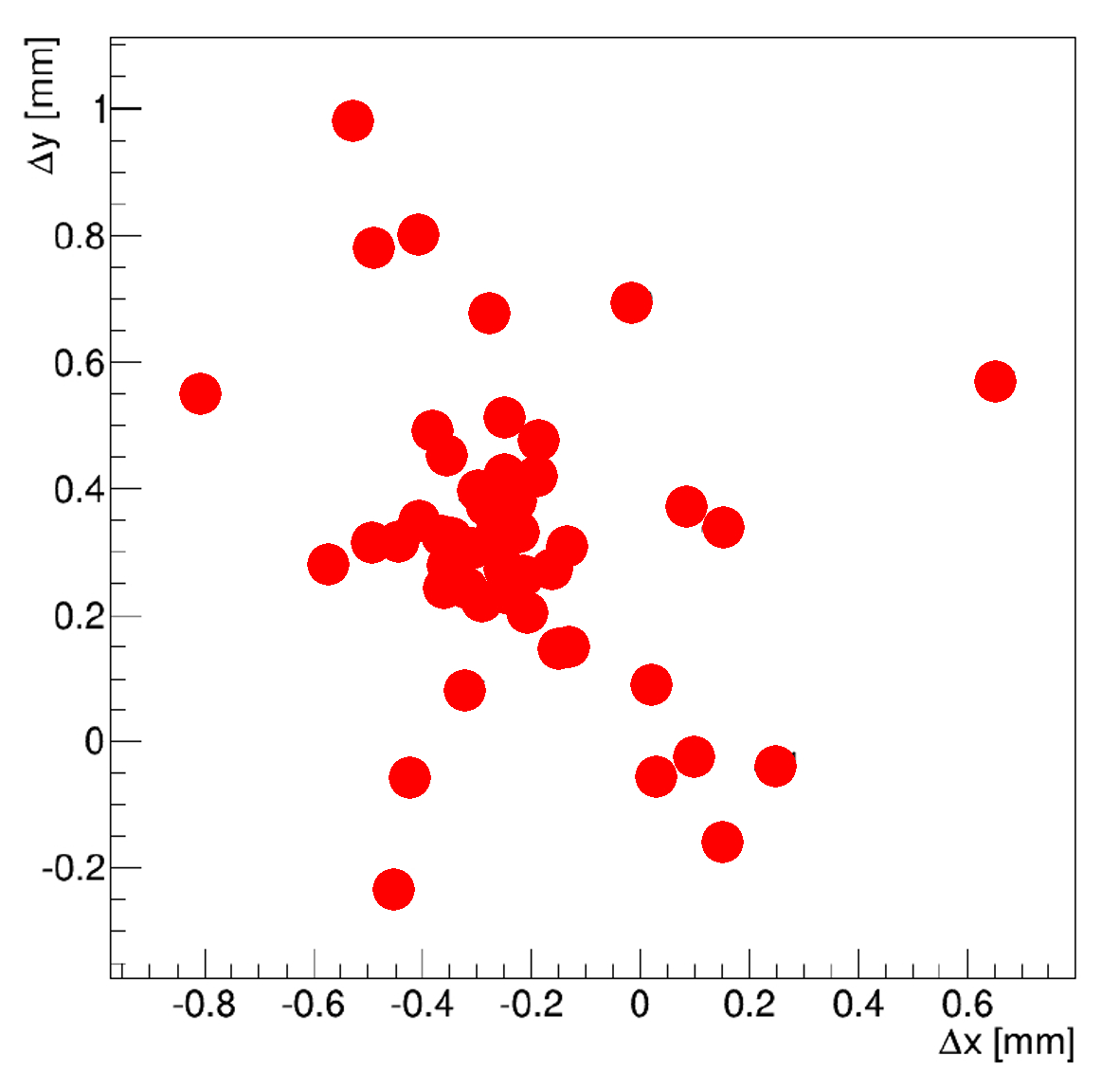}
   \caption{Misalignment of the MAPMT mask with respect to the fibre bundle for the 48 planes.}
   \label{fig:misalignment}
\end{figure}

The measurement of the crosstalk rate in the adjacent channels also provides a measurement of the misalignment of the MAPMT mask with respect to the fibre bundle.
If a mask is shifted, light is more likely to leak and create signals in the channel towards which it is offset. The centre of the mask with respect to the
centre of fibre bundle is computed as

\begin{equation}
(x_C,y_C)=\left(\frac{\sum_ix_iw_i}{\sum_iw_i},\frac{\sum_iy_iw_i}{\sum_iw_i}\right),
\end{equation}

with ($x_i,y_i$) the coordinates of the surrounding channels and $w_i$ the number of hits recorded in them. The resolution is a function of $1/\sqrt{N}$, where
$N$ is the amount of signals recorded, hence a high voltage is chosen for this analysis. The results for the 48 planes are presented in
figure~\ref{fig:misalignment}. There is a noticeable cluster around $(-0.3,0.3)$ but nothing that could impair the detector.

\subsection{Cosmic-based channel mismatch analysis}\label{sec:ch_mismatch}

The EMR design, involving the connection of external clear fibres to internal WLS fibres~\cite{emr_design_change}, leaves room for human error in
matching the two correctly. A dedicated analysis was developed in order to verify the consistency of this connection across the 2832 bars in the detector. 

This analysis~\cite{emr_xt, Francois} 
uses the distance between each bar hit and its particle track as a tool to estimate the likelihood of mismatch. A mismatched channel is
not reconstructed in the right location and is, on average, significantly less consistent with the other measurements in a reconstructed particle track.
Cosmic muons were particularly suited for this procedure as they cross the whole detector in straight lines, without stopping and provide full coverage.
When the data sample used in this analysis was recorded, the EMR was positioned up right with planes vertically oriented. Two cuts are applied to the data
sample in order to rid the muon tracks of artificial hits caused by crosstalk and noise. Crosstalk signals were rejected by placing a lower limit on the
ToT measurement (ToT$>$5). Restricting the delay between the trigger time and the hit time to a small interval was used to get rid of most of the noise
(-100$<\Delta t<$-80 ADC counts). 

To reconstruct tracks and calculate the distance of each hit from its particle trail, the hits were split into two projections $qz$, $q=x,y$. The plane ID
of the channel hit provides the $z$ coordinate and the bar ID provides either the $x$ or $y$ coordinates, depending on the plane orientation. The $(q_i,z_i)$
coordinates were those of the barycentre of the triangular section of the bar corresponding to the hit. For a linear fit $q=a_qz+b_q$, the absolute distance
between a hit $(q_i,z_i)$ and the track within a plane is $\Delta q_i=|q_i-(a_qz_i+b_q)|$. Distances are expressed in bar units ($\mathrm{b.u.}$) in the
following developments. A $\mathrm{b.u.}$ corresponds to the height of the triangular section or equivalently to the half width of its base.

The critical secondary variables, measured for each channel are the ratios of mismatch, $R_i$. Given an integer $i$, the ratio $R_i$ corresponds to the fraction
of the sample for which the bar is within $i\pm2/3$ $\mathrm{b.u.}$ off-track. For a distance distribution $f(\Delta)$, the ratio is defined as
\begin{equation}
R_i=\frac{\int_{i-\sigma}^{i+\sigma}f(\Delta)\mathrm{d}\Delta}{\int_{0}^{\sigma}f(\Delta)\mathrm{d}\Delta+\int_{i-\sigma}^{i+\sigma}f(\Delta)\mathrm{d}\Delta}.
\end{equation}
with $\sigma=2/3\,\mathrm{b.u.}$, an arbitrary overlapping uncertainty on the point.

For instance, the ratio $R_1$ represents the probability of a bar of being mismatched by exactly 1\,$\mathrm{b.u.}$, i.e. to be swapped with an adjacent bar.
It is shown in \cite{Francois} that $R_i$ is theoretically estimated to take values summarized in table~\ref{tab:mismatch_ratio} for different scenarios.
The $X-Y$ asymmetry of the $R_i$ ratio is due to the different angular distribution of muons seen by the vertical and the horizontal bars.

\begin{table}[!h]
 \centering
 \begin{tabular}{c|c|c|c|c}
  & \multicolumn{2}{c|}{Matched} & \multicolumn{2}{c}{Mismatched}  \\
  \hline
  & $xz$ proj. & $yz$ proj. & $xz$ proj. & $yz$ proj. \\
  \hline
  $R_1$ & 25.3\,\% & 32.2\,\% & 62.6\,\% & 66.1\%  \\
  $R_{i\geq2}$ & $\sim$0\,\% & $\sim$0\,\% & $\sim$100\,\% & $\sim$100\,\%
 \end{tabular}
 \caption{Mismatch ratios, $R_i$, for matched and mismatched bars in the two projections.}
 \label{tab:mismatch_ratio}
\end{table}

\begin{figure}[htr!]
  \begin{minipage}[b]{.49\textwidth}
   \centering
   \includegraphics[width=\textwidth]{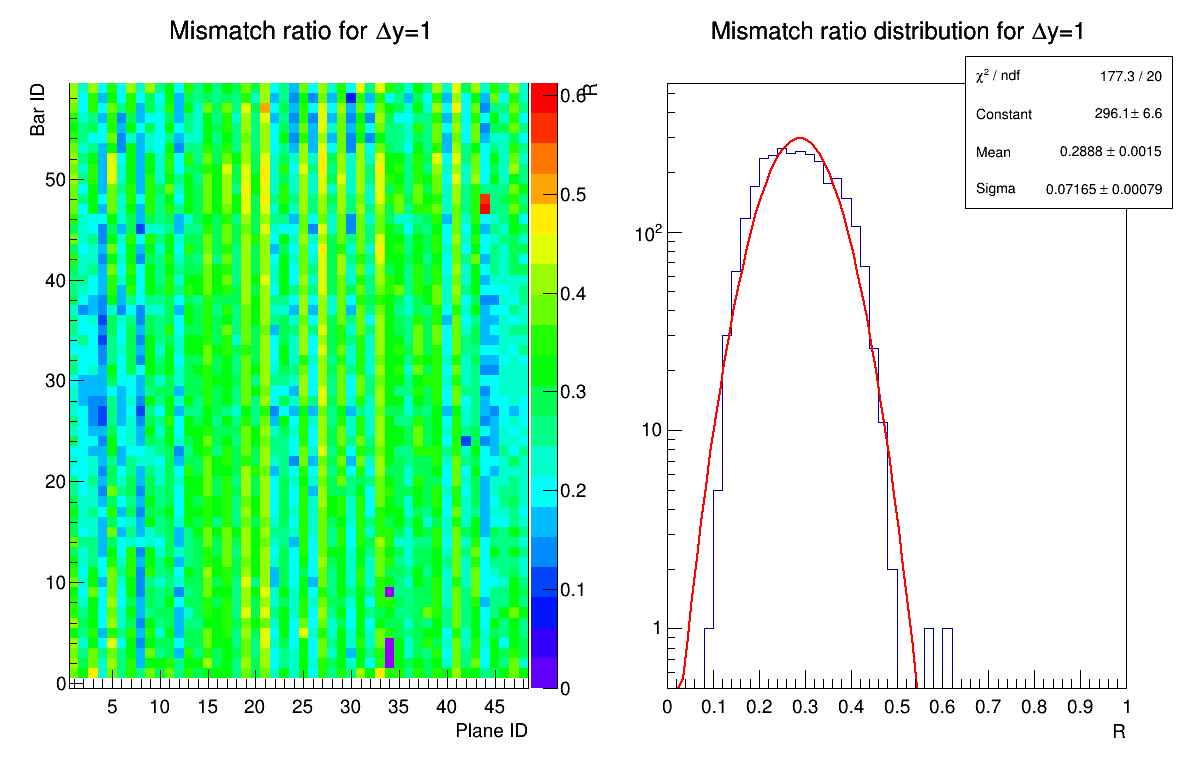}
   \caption{Mismatch ratio for adjacent bars.}
   \label{fig:ratio1}
  \end{minipage}
  \begin{minipage}[b]{.49\textwidth}
   \centering
   \includegraphics[width=\textwidth]{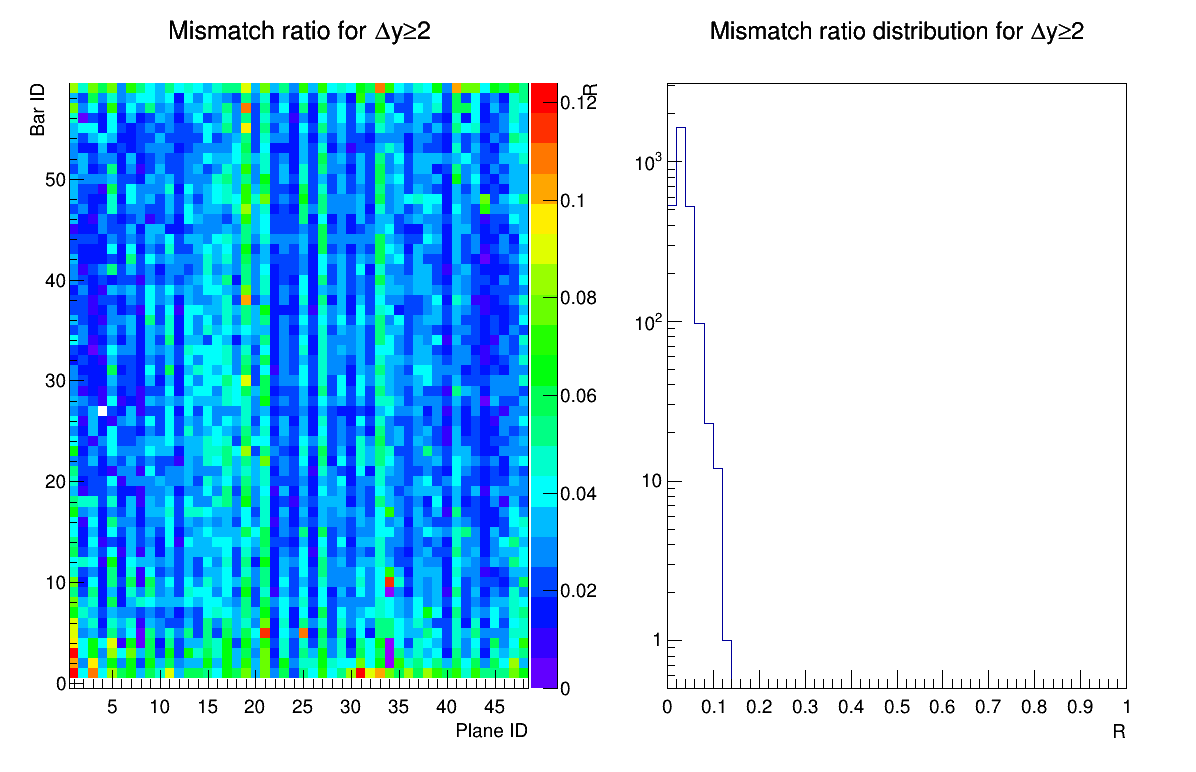}
   \caption{Mismatch ratio for distant bars.}
   \label{fig:ratio2}
  \end{minipage}
\end{figure}

The mismatch ratio for adjacent bars ($R_1$) was measured and is represented for each channel in figure \ref{fig:ratio1}, left. The ratio distribution is
represented next to it in log scale. The bulk of the distribution is centred around 28.88\,\%, consistent with the weighted average of the theoretical
predictions for the two projections. Bars 47 and 48 of plane 44 record a ratio of 62.5 $\pm$ 3.5\,\% and 57.2 $\pm$ 3.2\,\%, respectively, significantly
larger than the bulk one and in agreement with the prediction. Their proximity to each other corroborates the hypothesis of a channel swap. The mismatch
is fixed at the level of the channel map in the reconstruction software.

The same analysis has been performed for potential mismatches of two bars or more, $R_{i\geq2}$. Figure~\ref{fig:ratio2} represents the value of that ratio
for each channel. The results strongly reject any mismatch at this level.

\subsection{PMT Calibration}\label{sec:calib}
\begin{figure}[htb]
\centering
\includegraphics[width=\textwidth]{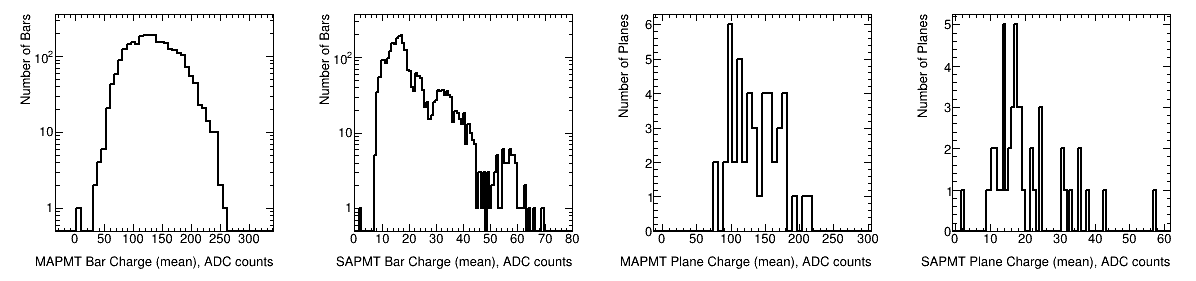}
\caption{Distributions of the first round of calibration. From left to right: mean MAPMT bar charge, mean SAPMT bar charge, mean MAPMT plane charge and mean
SAPMT plane charge.}
\label{fig:calib}
\end{figure}

A thorough charge calibration was performed shortly after completion of the detector. Cosmic data were used and the charge on the MAPMT side, $Q_M^i$ was
reconstructed by using equation~\ref{exp}. Provided these measurements, a calibration constant is produced for each MAPMT and SAPMT channel:
\begin{equation}
\epsilon_M^i=\frac{\overline{Q_M^i}}{\frac{1}{N}\sum_{j=1}^N\overline{Q_M^j}}\ , \hspace{10mm} \epsilon_S^i=\frac{\overline{Q_S^i}}{\frac{1}{N}\sum_{j=1}^N\overline{Q_S^j}},
\end{equation}
with $\overline{Q}$ the average charges measured in the channel and N the total number of channels under calibration. The indices $M$ and $S$ stand for MAPMT
and SAPMT, respectively. Provided calibration, each charge measured is corrected dividing it by its corresponding calibration constant. The measurements obtained
for the 5664 EMR channels are represented in figure~\ref{fig:calib}. The figure summarizes the mean charges recorded in the MAPMT and SAPMT for each channel and
each plane. These values are used to compute the calibration constants.

\section{Transportation and installation at Rutherford Lab}
The total weight of the EMR detector is 2.5 tonnes. Therefore a special care was taken to insure safety and shock-free transportation of the detector.
The detector was attached to special shock absorbers \cite{shockabsorbers} designed to withstand its weight and allow for shock absorption in all three
directions. The shock absorbers were then attached to a pallet by which the detector was handled. It was placed in a truck and delivered from the
University of Geneva to the Rutherford Appleton Laboratory (Didcot, Oxfordshire, UK) in September 2013.

Once delivered, the EMR detector was installed in the MICE hall and positioned vertically at the end of the existing MICE beamline. It was later exposed
to a beam whose parameters were varied in order to achieve different beam compositions and momenta. This data were used to verify the designed 
functionality of the detector, i.e. the ability to distinguish different particle types (muons, electrons and pions) and to measure their ranges
\cite{performance}. It was demonstrated that the detector is capable of identifying electrons with an efficiency of 98.6\%, providing a purity of the
MICE muon beam that exceeds 99.8\%.

\section{Conclusions}

This paper describes the design and the construction of the MICE Electron-Muon Range detector. The EMR is a fully-active scintillator detector, which
provides tracking and calorimetric information. A preliminary evaluation of its performance with cosmic 
and LED data is reported. The data-quality analysis revealed an excellent performs of the detector with only 0.2\,\% dead channels and cross-talk well
within the design requirements.

\section*{Acknowledgement}
The work described here was made possible by grants from Swiss National Science Foundation, in the framework of the SCOPES programme and the European
Community under the European Commission Framework Programme 7 (AIDA project, grant agreement no. 262025 and EuCARD-2, grant agreement no. 312453). We
gratefully acknowledge all sources of support.

\bibliographystyle{unsrt}
\bibliography{EMRConstr}

\begin{thebibliography}{10}

\bibitem{NFReference}
S.~Ozaki, R.~Palmer, M.~Zisman, and J.~Gallardo.
\newblock {Feasibility Study-II of a Muon-Based Neutrino Source}.
\newblock Technical report, 2001.
\newblock BNL-52623.

\bibitem{nf}
S.~Choubey and \emph{et al.}
\newblock {International Design Study for the Neutrino Factory, Interim Design
  Report}.
\newblock Technical report, 2011.

\bibitem{Neuffer:multitev}
D.~Neuffer.
\newblock {Multi-TeV Muon Colliders}.
\newblock {\em AIP Conf.Proc.}, 156:201--208, 1987.

\bibitem{Palmer:1996gs}
R.~Palmer, A.~Sessler, A.~Skrinsky, A.~Tollestrup, A.~Baltz, et~al.
\newblock {Muon collider design}.
\newblock {\em Nucl.Phys.Proc.Suppl.}, 51A:61--84, 1996.

\bibitem{icool1}
D.~Neuffer.
\newblock Principles and applications of muon cooling.
\newblock {\em Part. Accel.}, 14:75, 1983.

\bibitem{MICEweb}
Mice web site.
\newblock {\em http://mice.iit.edu}, contains detailed information about the
  experiment.

\bibitem{target}
C.~N.~Booth et~al.
\newblock The design, construction and performance of the {MICE} target.
\newblock {\em Journal of Instrumentation}, 8:P03006, 2013.

\bibitem{isis}
{ISIS} pulsed neutron and muon source at the {R}utherford~{A}ppleton
  laboratory.
\newblock web site.
\newblock {\em http://www.isis.stfc.ac.uk}.

\bibitem{TOFref}
R.~Bertoni and \emph{et al.}
\newblock {The design and commissioning of the MICE upstream time-of-flight
  system}.
\newblock {\em Nucl. Inst. Meth. A}, 615(1):14 -- 26, 2010.

\bibitem{MICE_PID}
M.~Rayner and M.~Bonesini.
\newblock {The MICE PID Detector System}.
\newblock Technical report, 2010.
\newblock http://mice.iit.edu/mnp/MICE0304.pdf, MICE-NOTE-304.

\bibitem{Bertoni:tof2}
R.~Bertoni, A.~Bonesini, A.~de~Bari, G.~Cecchet, Y~Karadzhov, and R.~Mazza.
\newblock {The construction of the {MICE} {TOF}2 detector}.
\newblock 2010.
\newblock http://mice.iit.edu/mnp/MICE0286.pdf, MICE-NOTE-DET-286.

\bibitem{ruslan}
R.~Asfandiyarov.
\newblock Totally active scintillator tracker-calorimeter for the {M}uon
  {I}onization {C}ooling {E}xperiment.
\newblock {\em University of Geneva, PhD thesis}, 2014.
\newblock http://dpnc.unige.ch/THESES/THESE\_ASFANDIYAROV.pdf.

\bibitem{Sandstrom:2007zz}
Anton~Rikard Sandstrom.
\newblock {\em {Background and Instrumentation in MICE}}.
\newblock PhD thesis, 2007.
\newblock https://inspirehep.net/record/776393/files/Thesis-2007-Sandstrom.pdf.

\bibitem{PlaDalmau:2001fr}
A.~Pla-Dalmau, A.~Bross, and K.~Mellott.
\newblock {Low-cost extruded plastic scintillator}.
\newblock {\em Nucl. Instrum. Meth.}, A466:482--491, 2001.
\newblock FERMILAB-PUB-00-177-E.

\bibitem{PlaDalmau:2005df}
A.~Pla-Dalmau, A.~Bross, V.~Rykalin, and B.~Wood.
\newblock {Extruded plastic scintillator for MINERvA}.
\newblock {\em Nuclear Science Symposium Conference Record, IEEE}, 3, 2005.
\newblock FERMILAB-CONF-05-506-E.

\bibitem{saintgobain}
Saint-Gobain Crystals.
\newblock Scintillating fiber brochure.

\bibitem{kuraray}
Kuraray.
\newblock Plastic scintillating fibers brochure.

\bibitem{emr_design_change}
R.~Asfandiyarov et~al.
\newblock Modifications to {EMR} design.
\newblock {\em MICE internal note}, 2011.
\newblock http://mice.iit.edu/mnp/MICE0357.pdf, MICE-NOTE-DET-357.

\bibitem{hamamatsu_mapmt}
Hamamatsu.
\newblock Multianode photo-multiplier tubes {R}5900-00-{M}64 datasheet.

\bibitem{harp}
M.~G.~Catanesi et~al. [HARP~Collaboration].
\newblock The {HARP} detector at the {CERN PS}.
\newblock {\em Nucl. Instr. and Meth.}, (A 571):527, 2007.

\bibitem{philips}
R.~Asfandiyarov et~al.
\newblock Selecting {P}hilips {XP} 2972 {P}hotomultiplier {T}ubes for the
  {E}lectron {M}uon {R}anger ({EMR}).
\newblock {\em MICE internal note}, 2012.
\newblock http://mice.iit.edu/mnp/MICE0383.pdf, MICE-NOTE-DET-383.

\bibitem{Bolognini2011108}
D.~Bolognini et~al.
\newblock Tests of the {MICE} {E}lectron {M}uon {R}anger frontend electronics
  with a small scale prototype.
\newblock {\em Nuclear Instruments and Methods in Physics Research Section A:
  Accelerators, Spectrometers, Detectors and Associated Equipment}, 646(1):108
  -- 117, 2011.

\bibitem{maroc}
S.~Franz and P.~Barrillon.
\newblock Atlas alfa-measuring absolute luminosity with scintillating fibres.
\newblock {\em Nuclear Instruments and Methods in Physics Research, Section A:
  Accelerators, Spectrometers, Detectors and Associated Equipment},
  610(1):35--40, 2009.

\bibitem{V1731}
CAEN.
\newblock V1731 4/8 ch. 8 bit 1000/500 ms/s {D}igitizer. {T}echnical
  information manual.

\bibitem{emr_elquality}
R.~Asfandiyarov, A.~Blondel, F.~Drielsma, and Y.~Karadzhov.
\newblock {E}lectron-{M}uon {R}anger ({EMR}) {E}lectronics {Q}uality {T}ests.
\newblock {\em MICE internal note}, 2014.
\newblock http://mice.iit.edu/micenotes/public/pdf/MICE0441/MICE0441.pdf,
  MICE-NOTE-DET-441.

\bibitem{emr_xt}
R.~Asfandiyarov, A.~Blondel, F.~Drielsma, and Y.~Karadzhov.
\newblock Crosstalk and {M}isalignment in the {E}lectron-{M}uon {R}anger
  ({EMR}).
\newblock {\em MICE internal note}, 2014.
\newblock http://mice.iit.edu/micenotes/public/pdf/MICE0440/MICE0440.pdf,
  MICE-NOTE-DET-440.

\bibitem{Francois}
F.~Drielsma.
\newblock Electron {M}uon {R}anger ({EMR}) hardware characterization.
\newblock {\em University of Geneva, Master thesis}, 2014.
\newblock http://mice.iit.edu/phd/EMR\_hardware\_characterization.pdf.

\bibitem{shockabsorbers}
Vibrationmounts.
\newblock http://vibrationmounts.com/RFQ/VM02019.htm.

\bibitem{performance}
D.~Adams et~al. MICE~collaboration.
\newblock Electron-muon ranger: performance in the {MICE} muon beam.
\newblock {\em Journal of Instrumentation}, 10:P12012, 2015.

\end{thebibliography}

\end{document}